\documentclass[showpacs,twocolumn]{revtex4}

\usepackage{graphics}
\usepackage{graphicx}

\def\m@thcombine#1#2{%
  \setbox0=\hbox{$#1$}
  \setbox1=\hbox{$#2$}
  \ifdim\wd0>\wd1
    \setbox0=\hbox to\wd1{\hss\box0\hss}
  \else
    \setbox1=\hbox to\wd0{\hss\box1\hss}
  \fi
  \mathop{\vcenter{
    \offinterlineskip\box0\box1}}}
\def\lesim{\m@thcombine<\sim}
\def\gesim{\m@thcombine>\sim}

\newcommand{\vecr}{\mbox{\boldmath $r$}}

\newcommand{\vecrs}{\mbox{\boldmath $r$}\sigma}

\newcommand{\vectau}{\mbox{\boldmath $\tau$}}

\newcommand{\rhot}{\tilde{\rho}}

\newcommand{\calA}{{\cal{A}}}
\newcommand{\calB}{{\cal{B}}}

\newcommand{\eps}{\epsilon}

\begin{document}

\title
{
Surface-enhanced pair transfer in 
quadrupole states of neutron-rich Sn isotopes
}

\author{Masayuki Matsuo, Yasuyoshi Serizawa}

\affiliation{
Department of Physics, Faculty of Science and 
Graduate School of Science and Technology, 
Niigata University, Niigata 950-2181, Japan 
}

\date{\today}

\begin{abstract}
We investigate the neutron pair transfer modes associated with the low-lying quadrupole states 
in neutron-rich Sn isotopes
by means of the quasiparticle random phase approximation based on the 
Skyrme-Hartree-Fock-Bogoliubov mean field model. The transition strength of the quadrupole
pair-addition mode feeding the $2_1^+$ state is enhanced
in the Sn isotopes with $A \geq 132$. 
The transition density of the pair-addition mode has a large spatial 
extension in the exterior of nucleus, reaching far to $r\sim 12-13$ fm.
The quadrupole pair-addition mode reflects sensitively  
a possible increase of the effective pairing interaction strength in the surface and
exterior regions of 
neutron-rich nuclei.
\end{abstract}


\maketitle

\section{Introduction}\label{Intro}

The pair correlation is one of the fundamental nucleon many-body
correlations, which governs 
many aspects of the nuclear structure\cite{BM2,Brink-Broglia,Shimizu89,Dean03}. 
Recently attentions are paid to the properties of the
pair correlation in neutron-rich nuclei, which exhibit
new features such as the neutron halo, the neutron skin
and the presence of weakly bound neutrons. 
For example, 
a spatial two-neutron correlation or the di-neutron correlation,  
between two weakly-bound neutrons forming the halo in
$^{11}$Li and $^{6}$He has been discussed extensively
\cite{Esbensen,Ikeda,Zhukov,Hagino05,Hagino07,Kikuchi09},
and its experimental signatures are reported recently
\cite{Nakamura,Mueller-He}. 
There exists also several theoretical studies demonstrating 
similar spatial two-neutron correlation 
in many other nuclear systems. The spatial two-particle correlation
was originally discussed in the closed-shell + two-particle systems
\cite{Ibarra,Janouch,Catara84,Ferreira}, and recent selfconsistent
Hartree-Fock-Bogoliubov
mean-field calculations suggest that 
the correlation is generically enhanced 
around the nuclear surface of neutron-rich 
nuclei covering also medium- and heavy-mass regions
\cite{MMS05,Pillet07,Pillet10}.
Furthermore it is found that
neutron matter at low densities may exhibit features of 
the strong coupling pairing\cite{Leggett1,Leggett2,Nozieres},
characterized by the small size ($\sim 5$ fm at minimum) of 
the Cooper pair comparable with the average inter-neutron distance
\cite{Matsuo06,Margueron08,AbeSeki09},
reflecting the large scattering length $a=-18.5$fm of the bare nuclear force.
This may also suggest a possible enhancement of the pair correlation in 
the surface and exterior regions of neutron-rich nuclei.
The spatial two-neutron correlation and the surface enhancement of the 
neutron pairing is
also discussed in the cases of a slab and a semi-infinite nuclear 
matter\cite{Enyo09,Pankratov09}. Also the pairing interaction induced by
the coupling to the phonons is expected to increase the 
surface enhancement\cite{Pastore}.

It is important to ask how one can probe the surface enhancement of the 
pair correlation in neutron-rich nuclei. We consider here the transfer
of a neutron pair since the matrix element of the pair
transfer has direct relation to the degree of the pair
correlation\cite{Yoshida62,Bes-Broglia66,Broglia73,Brink-Broglia,BM2}. 
Recently the neutron pair transfer process in light neutron-rich 
nuclei $^{11}$Li and $^{6,8}$He has been studied experimentally
\cite{Tanihata08,Chatterjee08,Keeley07,Golovkov09}.
In the theoretical side,
a possible link between the pair transfer between the $0^+$ ground states
and the surface enhancement of
the pairing in medium and heavy neutron-rich nuclei was suggested by 
Dobaczewski {\it et al.}\cite{DobHFB2}.
Recently the pair transfer feeding
the excited $0^+$ states in neutron-rich nuclei has been
investigated
\cite{Khan04,Avez,Khan09} 
from the viewpoint of the
pair vibrational mode, which are collective vibrational
modes associated with the pair correlation\cite{Bes-Broglia66,BM2,Brink-Broglia}.
Khan {\it et al.}\cite{Khan09}
has argued that the pair transfer can be a possible probe of
different models of the pairing interactions
with differences in density dependence and surface enhancement.

In the present work, we shed light on the neutron pair transfer with the {\it quadrupole}
multipolarity\cite{Bes-Broglia71,Broglia-Bes-Nilsson74,Broglia73},i.e.,
the transitions from the ground state of
a nucleus with neutron number $N$ feeding excited $2^+$ states, in particular
the first $2^+$ state, in the neighboring
$N\pm 2$ isotopes. 
It is noted that 
the first $2^+$ state in open shell
nuclei is a collective state of the surface vibrational type
with a strong influence by the pair correlation\cite{BM2}. With this surface
character of the mode,
we expect that the pair transfer feeding the $2^+_1$ state may provide information
on the pair correlation possibly enhanced 
in the surface and exterior regions of neutron-rich nuclei.

The low-lying quadrupole vibration is revealed recently to show new features in some
neutron-rich nuclei. For instance, the ratio between the neutron and
proton amplitude becomes significantly different from the liquid drop
ratio $N/Z$\cite{Ong06,Ong08,Elekes09,Jewell99,Khan00,Iwasa08}. 
New aspects of the quadrupole vibration may emerge also in 
the pair transfer transitions feeding 
the $2^+$ states.  
In our previous studies on the dipole and octupole responses, we have shown 
that soft modes accompanying a large pair transition amplitude in the surface and
exterior regions
emerge above the 
threshold energy in nuclei with small neutron separation 
energy\cite{MMS05,Serizawa09,Matsuo07}. 
We also found that these soft dipole and octupole modes are sensitive to the 
surface-enhancement of the pair correlation.
The low-lying quadrupole vibration is
different from these soft modes in many respects: 
it exists systematically
from stable nuclei
to nuclei far from the stability line, and the excitation energy is located 
below the separation energy in most of the cases.  
Nevertheless, as we shall show below, 
the low-lying quadrupole mode also has sensitivity to 
the surface pair correlation. The sensitivity is systematically
seen in the pair-addition transfer feeding the lowest-lying quadrupole state,
 and it becomes significant in neutron-rich nuclei
beyond the shell closure, i.e. in  Sn isotopes with $A \geq 132$.

We shall perform systematic numerical study by adopting
the Sn isotopes as a representative case since its long isotopic chain enables us to
see systematic trends. In particular we focus on 
isotopes beyond the $N=82$ shell gap, above which the neutron separation
energy  suddenly decreases. 
Our study is based on the Skyrme-Hartree-Fock-Bogoliubov
mean-field theory in describing the pair correlated ground state,
and the quasiparticle random phase approximation (QRPA)\cite{Serizawa09} (and references
in \cite{PVKC07}) 
to describe the
$2^+$ states and the associated pair transfer modes. In the
present work, we concentrate on the nuclear structure aspects, and leave 
the cross section and
the reaction mechanism of the neutron pair transfer\cite{Broglia73,Igarashi91,
Oertzen-Vitturi,Potel} for future works. 
The formalism is described in
Section 2. To elucidate effects of the surface-enhanced pairing,
we compare
different models of the effective pairing interaction, which have
and do not have the property of the surface enhancement. 
After presenting the adopted effective pairing interactions in
Section 3,
we discuss numerical
results in Section 4, and the conclusions are drawn in Section 5.

\section{Skyrme-Hartree-Fock-Bogoliubov mean-field plus QRPA approach}

\subsection{model}

We adopt the Hartree-Fock-Bogoliubov mean-field model 
and the quasiparticle random phase approximation (QRPA) to describe the
pair transfer modes with the quadrupole multipolarity. 
The basic framework of the present HFB+QRPA formalism is described in 
Refs.\cite{Matsuo01,Serizawa09}.
Here we recapitulate it briefly in order to explain the procedure 
to describe the 
pair transfer modes.

In performing the HFB calculation, we assume the spherical symmetry for 
the ground state  and the associated mean-fields.
The Skyrme energy density functional (the energy density functional associated with the Skyrme interaction)
is adopted to derive 
the Hartree-Fock potential
and the position-dependent effective mass\cite{Bender}. The parameter set SLy4 is adopted\cite{CBH98}.
As the residual interaction responsible for the pairing correlation we  use
the density-dependent delta interaction (DDDI) \cite{DobHFB2,
DDpair-Chas,Esbensen,Garrido,Garrido01,
DD-Dob,DD-mix1,DD-mix2,Hagino05,Matsuo06,Yam05,Yam09}
\begin{equation}\label{DDDI}
v^{{\rm pair}}_q(\vecr,\vecr')={1\over2}(1-P_\sigma)V_q(\vecr)
\delta(\vecr-\vecr') \ \ \ (q=n,p)
%\left(1-\eta{\rho(\vecr)\over \rho_0}\right)\delta(\vecr-\vecr')
\end{equation}
where the the pairing interaction strength $V_q(\vecr)$ is a function of the
neutron and proton densities  $\rho_q(\vecr)$ ($q=n,p$), and hence
it depends on the position $\vecr$. 
We solve
the Hartree-Fock-Bogoliubov equation in the radial coordinate space 
to obtain the wave functions of the quasiparticle states. 
We assume the
box boundary condition at $r=R_{{\rm max}}$ with $R_{{\rm max}}=20$ fm. 
We truncate 
the quasiparticle states by the maximum quasiparticle  energy 
 $E_{\rm max}=60$ MeV. We also truncate the quasiparticle states
 by the maximum orbital angular quantum number $l_{\rm max}=12$, which is
 sufficiently large to obtain the converged HFB solutions of 
the Sn isotopes.

The QRPA calculation is then performed on the basis of the self-consistent HFB mean-fields.
As the residual interaction
in the particle-particle channel, we adopt the DDDI, Eq.~(\ref{DDDI}),  
the same effective pairing interaction
used for the HFB calculation.  Concerning the residual interaction in 
the particle-hole channel, we derive it also from the Skyrme functional used for the 
HFB calculation,
but we apply the Landau-Migdal approximation to the residual
interaction\cite{Khan04,Khan09,PVKC07,Serizawa09}. 
Namely, 
we approximate the particle-hole residual interaction by a position-dependent
contact interaction
$v_{ph}(\vecr,\vecr')= 
\left\{(F_0/N_0)(\vecr) + (F'_0/N_0)(\vecr)\vectau\cdot\vectau') \right\}
\delta(\vecr-\vecr')$
where $F_0/N_0$ and $F'_0/N_0$ are the Landau-Migdal parameters
derived from the Skyrme functional. The position dependence is 
taken into account 
in the Landau parameters 
by the local density approximation to the Fermi momentum.
As a prescription  commonly adopted in the QRPA studies using the Landau-Migdal
approximation\cite{Khan04,Khan09,PVKC07,Serizawa09}, we introduce a
renormalization factor $f$ multiplying $v_{ph}$, and it is fixed to impose the
zero excitation energy for the lowest-energy isoscalar dipole solution corresponding
to the spurious center-of-mass motion.

We adopt the linear response formalism for the QRPA calculation\cite{Matsuo01}.
Given the above residual interactions, we have coupled linear response equations for
three kinds of fluctuating densities $\delta\rho_q(\vecr,\omega)$,
$\delta\rhot_{q,+}(\vecr,\omega)$ and $\delta\rhot_{q,-}(\vecr,\omega)$,
which correspond respectively to fluctuating parts of the nucleon density
\begin{eqnarray}\label{rhofl}
\rho_q(\vecr,t) &\equiv&
\langle \Psi(t)|\hat{\rho}_q(\vecr)|\Psi(t)\rangle \nonumber \\
&=& 
\langle \Psi(t)|\sum_\sigma\psi_q^\dagger(\vecr\sigma)\psi_q(\vecr\sigma)|\Psi(t)\rangle
\end{eqnarray}
and of two kinds of the pair densities 
\begin{equation}\label{pairpm}
\rhot_{q,\pm}(\vecr,t) 
\equiv
\langle \Psi(t)|
\psi_q^\dagger(\vecr\downarrow)\psi_q^\dagger(\vecr\uparrow) 
\pm \psi_q(\vecr\uparrow)\psi_q(\vecr\downarrow) 
|\Psi(t)\rangle.
\end{equation}
The time-dependence of the state vector, and hence 
the density fluctuations 
$\delta\rho_q(\vecr,\omega)=Y_{LM}(\hat{\vecr})\delta\rho_q(r,\omega)/r^2$ and
$\delta\rhot_{q,\pm}(\vecr,\omega)=Y_{LM}(\hat{\vecr})\delta\rhot_{q,\pm}(r,\omega)/r^2$
in the frequency domain arise from
a response of the system caused by external fields with multipolarity $L$.
It is convenient to introduce operators 
\begin{eqnarray}
P_q^\dagger(\vecr)&=&\frac{1}{2}\sum_\sigma
\psi_q^\dagger(\vecr\sigma)\psi_q^\dagger(\vecr\tilde{\sigma}),\\
P_q(\vecr)&=&\frac{1}{2}\sum_\sigma
\psi_q(\vecr\tilde{\sigma}) \psi_q(\vecr\sigma).
\end{eqnarray}
Here  $P_q^\dagger(\vecr)$ and $P_q(\vecr)$ represent respectively
addition and removal at position $\vecr$ of
a pair of like-nucleons whose spins are coupled to the singlet $S=0$.
The pair densities are written compactly as
$\rhot_{q,\pm}(\vecr,t) 
=
\langle \Psi(t)|P_q^\dagger(\vecr)\pm P_q(\vecr) |\Psi(t)\rangle. $

For a response with the multipolarity $L$ and the
natural parity $\pi=(-)^L$, the coupled
linear response equation is written in the radial coordinate system as
\begin{widetext}
\begin{equation} \label{rpa}
\left(
\begin{array}{c}
\delta\rho_{q L}(r,\omega) \\
\delta\rhot_{+,q L}(r,\omega) \\
\delta\rhot_{-,q L}(r,\omega) 
\end{array}
\right)
=\int_0^{R_{{\rm max}}} dr'
\left(
\begin{array}{ccc}
& & \\
& R_{0,q L}^{\alpha\beta}(r,r',\omega)& \\
& & 
\end{array}
\right)
\left(
\begin{array}{l}
\sum_{q'}\kappa_{ph}^{qq'}(r')
\delta\rho_{q' L}(r',\omega)/r'^2 + v^{{\rm ext}}_{0, q L}(r') \\
\kappa_{{\rm pair}}(r')\delta\rhot_{+,q L}(r',\omega)/r'^2 + \tilde{v}^{{\rm ext}}_{+, q L}(r') \\
-\kappa_{{\rm pair}}(r')\delta\rhot_{-,q L}(r',\omega)/r'^2 + \tilde{v}^{{\rm ext}}_{-, q L}(r')
\end{array}
\right),
\end{equation}
\end{widetext}
where $R_{0,qL}^{\alpha\beta}$ is the unperturbed response function
representing non-interacting two-quasiparticle
excitations. $\kappa_{ph}(r)$ and $\kappa_{{\rm pair}}(r)$ are the residual
interactions in the particle-hole and the particle-particle channels, respectively.

In the present studies we consider
external fields of the pair addition and removal types
\begin{eqnarray}
V^{{\rm ext}}_{{\rm pair}}=&\int d\vecr \tilde{v}^{{\rm ext}}_{+, q L}(r)Y_{LM}(\hat{\vecr}) 
(P_q^\dagger(\vecr)+ P_q(\vecr)) \nonumber \\
&+\int d\vecr \tilde{v}^{{\rm ext}}_{-, q L}(r)Y_{LM}(\hat{\vecr}) 
(P_q^\dagger(\vecr)- P_q(\vecr))
\end{eqnarray}
in addition to 
the usual external field of the particle-hole type
\begin{equation}
V^{{\rm ext}}_{ph}=\int d\vecr \sum_q v^{{\rm ext}}_{0, q L}(r)Y_{LM}(\hat{\vecr}) \hat{\rho}_q(\vecr).
\end{equation}
Note that the linear response equation (\ref{rpa}) formulated
in the present work contains 
$\tilde{v}^{{\rm ext}}_{\pm, q L}(r)$ representing the pair-addition/removal fields.

For the unperturbed response function $R_{0,q L}^{\alpha\beta}$
appearing in the linear response equation (\ref{rpa}),
we adopt the spectral representation
\begin{widetext}
\begin{eqnarray}\label{usresp}
R_{0,q L}^{\alpha\beta}(r,r',\omega) =
{1\over 2}  \sum_{nlj,n'l'j'} &  
{\left<l'j'\right\|Y_L\left\|lj\right>^2 \over 2L+1} \hspace{70mm} \nonumber\\
& \left\{ 
\frac{\left<0|\calA|nlj,n'l'j'\right>_r
\left<nlj,n'l'j'|\calB|0\right>_{r'}}
{\hbar\omega + i\eps -E_{nlj}-E_{n'l'j'}}
 -
\frac{\left<0|\calB|nlj,n'l'j'\right>_{r'}
\left<nlj,n'l'j'|\calA|0\right>_{r}}
{\hbar\omega + i\eps +E_{nlj}+E_{n'l'j'}}
 \right\},
\end{eqnarray}
\end{widetext}
with
\begin{eqnarray}
\left<0|\calA|nlj,n'l'j'\right>_r & \equiv
\bar{\phi}_{n'l'j'}^{\rm T}(r) \calA \phi_{nlj}(r), \nonumber\\
\left<nlj,n'l'j'|\calA|0\right>_{r} & \equiv 
\phi_{nlj}^{\rm T}(r) \calA \bar{\phi}_{n'l'j'}(r), \nonumber
\end{eqnarray}
using 
the discretized quasiparticle eigenstates of the HFB mean-field Hamiltonian. 
Here $\phi_{nlj}(r)=(\phi_{nlj}^{(1)}(r), \phi_{nlj}^{(2)}(r))^{T}$ is the 
two-component radial wave function of a quasiparticle state with the quantum
number $nlj$, and $\calA$ and $\calB$ are $2\times 2$ matrices defined
for the three kinds of density $\rho_q$ and  $\tilde{\rho}_{q,\pm}$, 
Eqs.(\ref{rhofl}) and (\ref{pairpm}) (see Ref.\cite{Matsuo01}, for details).
$\epsilon$ is a smearing parameter, for which in the present work we adopt
$\epsilon=0.05$ MeV corresponding to a convolution of
the strength function with the Lorentzian function of FWHM=100 keV.
In the previous works\cite{Matsuo01,MMS05,Matsuo07,Serizawa09} we used the formalism which implements the continuum
quasiparticle states satisfying the proper
boundary conditions connecting to out-going waves. The continuum QRPA calculation
based on this formalism is useful for the states above the particle threshold energy, 
but, on the other hand, it costs quite severely a large 
computation time, especially
when we choose a small value for the smearing parameter $\epsilon$, and it is too demanding 
for systematic studies. The calculated lowest $2^+$ mode obtained in the present study 
are all located below the neutron separation energy, and we use a large box
size $R_{{\rm max}}=20$ fm.
In the present study, therefore, we adopt the spectrum representation using
the discretized 
quasiparticle states obtained in the HFB calculation. An
argumentation on this point will be given later.

\subsection{Strength function and transition density for
 pair transfer modes}

The HFB+QRPA model enables us to calculate the transition matrix elements
for pair transfer operators.  Here by the pair transfer operator
we mean a generalized 
one-body operator which either add or remove a pair of singlet neutrons.
We consider the following pair-addition operator with multipolarity $L$,
\begin{eqnarray}
P^\dagger_{LM}&=& \int d\vecr Y_{LM}(\hat{r})f(r)P^\dagger(\vecr) \nonumber\\
&=& \int d\vecr Y_{LM}(\hat{r})f(r)
\psi^\dagger(\vecr\downarrow)\psi^\dagger(\vecr\uparrow),
\end{eqnarray}
which is essentially the same as the multipole pair operators used in
the traditional works\cite{Bes-Broglia66,Bes-Broglia71,Broglia-Bes-Nilsson74,Broglia73}.
Here and hereafter we omit the isospin index $q$ for
simplicity  since  we discuss only the neutron pair
transfer and the neutron pairing in the present paper. 
Note that $f(r)$ is a form factor of the pair-addition operator,
and we choose $f(r)=1$
in the present analysis.
This is a simplification, but even this simple choice is useful 
for discussions   
on isotopic trends and roles of the surface enhanced 
pair interactions as we show below.

The strength function for the multipole pair-addition operator $P^\dagger_{LM}$ is
given by 
\begin{equation}
S_{{\rm Pad}L}(E)\equiv \sum_{iM} \delta(E-E_{iL}) |\langle \Psi_{iLM} |P^\dagger_{LM}|\Psi_0\rangle|^2,
\end{equation}
where $\Psi_0$ and $\Psi_{iLM}$ are the ground and the excited states obtained in 
the QRPA, and $E_{iL}$ is the QRPA excitation energy.
In the linear response formalism of QRPA, this quantity
can be calculated as 
\begin{equation}
S_{{\rm Pad}L}(\hbar\omega)=-\frac{2L+1}{\pi}{\rm Im}\int dr \frac{1}{2}(
\delta\rhot_{+,q L}(r,\omega) - \delta\rhot_{-,q L}(r,\omega)) 
\end{equation}
using the solutions of Eq.(\ref{rpa}) obtained with the external field  
\begin{equation}
\mathcal{V}^{{\rm ext}}(r)\equiv
\left(
\begin{array}{c}
v^{{\rm ext}}_{0, q L}(r)  \cr
\tilde{v}^{{\rm ext}}_{+, q L}(r)  \cr
\tilde{v}^{{\rm ext}}_{-, q L}(r)  
\end{array}
\right)
=
\left(
\begin{array}{c}
0  \cr
\frac{1}{2}  \cr
\frac{1}{2}
\end{array}
\right),
\end{equation}
which corresponds to an external field identified to the pair-addition operator,
i.e., $V^{{\rm ext}}_{{\rm pair}}=P^\dagger_{L0}$. We evaluate the strength
\begin{eqnarray}
B({\rm Pad}L)&=&\sum_{M} |\langle \Psi_{iLM}
 |P^\dagger_{LM}|\Psi_0\rangle|^2  \nonumber\\
&=&(2L+1) |\langle \Psi_{iL0} |P^\dagger_{L0}|\Psi_0\rangle|^2
\end{eqnarray}
associated with an individual QRPA solution
by integrating the strength function $S_{{\rm Pad}L}(E)$ in an energy interval
$E \in [E_i - 10\eps, E_i+10\eps]$, where $E_i$ is the peak energy 
of a peak in 
$S_{{\rm Pad}L}(E)$, corresponding to a QRPA solution, and $\eps$
is the smearing parameter.
An isolated discrete peak has the FWHM of $2\eps=100$ keV.

Similarly, we can consider the multipole pair-removal operator 
\begin{equation}
P_{LM}
= \int d\vecr Y^*_{LM}(\hat{r})f(r)
\psi_q(\vecr\uparrow)\psi_q(\vecr\downarrow)= (P^\dagger_{LM})^\dagger,
\end{equation}
the associated strength function
\begin{equation}
S_{{\rm Prm}L}(E)\equiv \sum_{iM} \delta(E-E_{iL}) |\langle \Psi_{iL-M} |P_{LM}|\Psi_0\rangle|^2,
\end{equation}
and  the strength of the pair-removal operator for the individual QRPA solution
\begin{eqnarray}
B({\rm Prm}L)&=&\sum_{M} |\langle \Psi_{iL-M} |P_{LM}|\Psi_0\rangle|^2 
\nonumber\\
&=&(2L+1) |\langle \Psi_{iL0} |P_{L0}|\Psi_0\rangle|^2.
\end{eqnarray}
The former is calculated as
\begin{equation}
S_{{\rm Prm}L}(\hbar\omega)=-\frac{2L+1}{\pi}{\rm Im}\int dr \frac{1}{2}(
\delta\rhot_{+,q L}(r,\omega) + \delta\rhot_{-,q L}(r,\omega)) 
\end{equation}
under the external field of the multipole pair-removal operator $P_{L0}$:  
\begin{equation}
\mathcal{V}^{{\rm ext}}(r)
=
\left(
\begin{array}{c}
0  \cr
\frac{1}{2}  \cr
-\frac{1}{2}
\end{array}
\right).
\end{equation}
When we calculate the strength 
$B({\rm Q}L)=(2L+1)|\langle \Psi_{iL0}|Q_{L0}|\Psi_0\rangle|^2$
for the multipole operator of the particle-hole type
$Q_{LM}=\int d\vecr Y_{LM}(\hat{r})r^L \hat{\rho}_q(\vecr)$,
we choose the external field as
$\mathcal{V}^{{\rm ext}}(r)
=\left(r^L, 0, 0 \right)^T$.

We also calculate the transition density
associated with the pair-addition  
$\Psi_{0} \stackrel{P^\dagger}{\longrightarrow} \Psi_{iLM}$
and pair-removal transfer
$\Psi_{0} \stackrel{P}{\longrightarrow} \Psi_{iL-M}$.
The transition density
associated with the addition/removal of a neutron
singlet pair at the position $\vecr$ 
is defined and factorized as
\begin{eqnarray}
P_{iLM}^{({\rm ad})}(\vecr)&\equiv&
\langle \Psi_{iLM} |
\psi^\dagger(\vecr\downarrow)\psi^\dagger(\vecr\uparrow)| \Psi_0\rangle  
\nonumber\\
&=&
\langle \Psi_{iLM} | P^\dagger(\vecr)| \Psi_0\rangle 
=Y_{LM}^*(\hat{\vecr})P_{iL}^{({\rm ad})}(r), \\
P_{iLM}^{({\rm rm})}(\vecr) &\equiv&
\langle \Psi_{iLM} | \psi(\vecr\uparrow)\psi(\vecr\downarrow)| \Psi_0\rangle 
\nonumber\\
&=&
\langle \Psi_{iLM} | P(\vecr)| \Psi_0\rangle 
=Y_{LM}^*(\hat{\vecr})P_{iL}^{({\rm rm})}(r).
\end{eqnarray}
Here the radial transition densities are 
expressed using the solution of the linear response equation as
\begin{eqnarray}
P^{({\rm ad})}_{iqL}(r)&=&{C\over 2\pi r^2}{\rm Im}
(\delta\rhot_{+,qL}(r,\omega_i) -\delta\rhot_{-,qL}(r,\omega_i) ), \\
P^{({\rm rm})}_{iqL}(r)&=&{C\over 2\pi r^2}{\rm Im}
(\delta\rhot_{+,qL}(r,\omega_i) +\delta\rhot_{-,qL}(r,\omega_i) )
\end{eqnarray}
where $E_i=\hbar\omega_i$ is the excitation energy of the QRPA solution,
and $C$ is a normalization constant. Note that the 
conventional transition density of the particle-hole type is given 
by
\begin{eqnarray}
\rho^{ph}_{iq}(\vecr)&\equiv &
\langle \Psi_{iLM}| \sum_\sigma\psi^\dag_q(\vecrs)
\psi_q(\vecrs) | \Psi_0\rangle \nonumber\\ 
&=&
Y_{LM}^*(\hat{\vecr})\rho^{ph}_{iqL}(r),  \\ 
\rho^{ph}_{iqL}(r)&= & 
-{C\over \pi r^2}{\rm Im} \delta\rho_{qL}(r,\omega_i).
\end{eqnarray}
The normalization factor $C$ is fixed so that it is
consistent with the strengths $B({\rm Q}L)$ or 
$B({\rm Pad}L)$ of the QRPA solution under consideration.

\begin{figure}[htb]
\includegraphics[scale=0.5,angle=270]{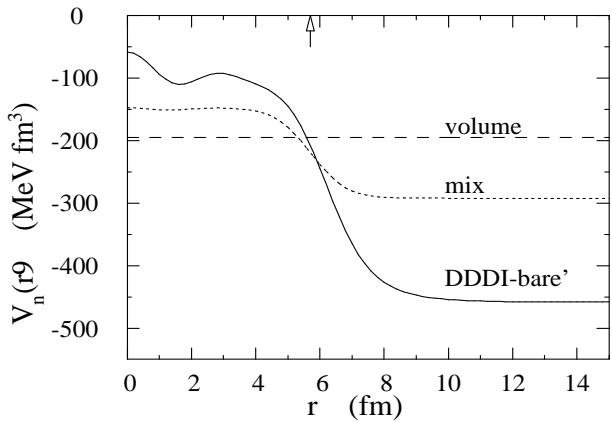}
\caption{\label{fig:DDDI} 
The pairing interaction strength $V_n(r)$ 
for neutrons in $^{134}$Sn, plotted as a 
function of the radial coordinate $r$
for the DDDI-bare', the 
mix and the volume pairing interactions. The position of the half density
radius is marked with the arrow.
}
\end{figure}

\begin{figure}[htb]
\includegraphics[scale=0.5,angle=270]{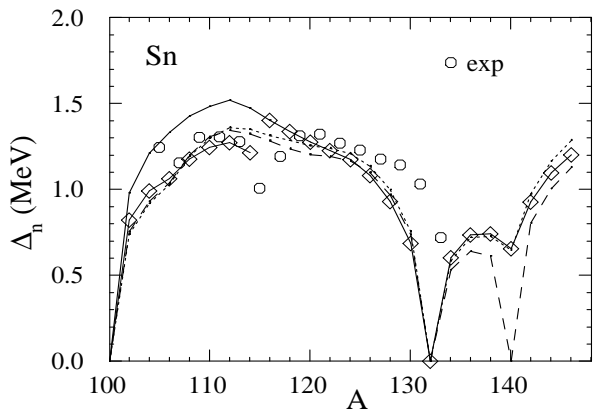}
\caption{\label{fig:gap} 
The average neutron pairing gap $\Delta_{\rm n}$ in the
Sn isotopes. The one calculated with the DDDI-bare' ($\eta=0.71$)
is plotted with the solid line. For $100 \leq A \leq 114$,
results obtained with the DDDI-bare' but using $\eta=0.75$ is
also plotted with the thin solid line and the diamond symbols. 
The ones calculated with the mix pairing and the volume pairing
are plotted with the dotted and the dashed lines, respectively.
The experimental neutron gap estimated using
the three-point formula  of the odd-even mass difference\cite{SDN98} 
and the mass compilation\cite{AME03} 
is also plotted with the circles. 
}
\end{figure}

\section{Effective pairing interactions}

In the present paper we intend to disclose influences of the surface
enhanced pairing on the low-lying quadrupole vibrational mode. 
For this purpose, we compare three 
versions of DDDI, Eq.(\ref{rpa}), 
which have different properties concerning the surface enhancement. They are
1) the density-dependent pairing interaction which has a connection to the bare
nuclear force via the scattering length,
2) the density-independent delta interaction,
and 3) the so-called mix pairing interaction\cite{DD-Dob,DD-mix1,DD-mix2}.

1) The first is the one which we adopt mainly in the following
analysis, and it is a version of DDDI which reproduces
properties of the bare nuclear force relevant for the pairing, in particular, 
the scattering length of the bare force\cite{Esbensen,Garrido,Garrido01,Matsuo06,Margueron08}. 
Among such DDDI's,
we adopt the parametrization
\begin{equation}
V_n(\vecr)=v_0 \left(1-\eta\left(\frac{\rho_n(\vecr)}{\rho_c}\right)^\alpha\right)
\end{equation} 
for the interaction of neutrons\cite{Matsuo06,Matsuo07}. Here the parameter $v_0$ is 
chosen to reproduce the scattering length $a=-18.5$fm in 
the $^{1}S$ channel.
The density-dependence characterized by the power $\alpha$ and
and the prefactor $\eta$ is adjusted to reproduce the neutron pairing
gap in dilute and uniform matter, more specifically the pairing 
gap obtained with the 
bare force and the BCS approximation. We find a parameter set  
$v_0=-458.4$ MeV fm$^{-3}$, $\rho_c=0.08$fm$^{-3}$, $\alpha=0.845$, and 
$\eta=0.845$ for the quasiparticle energy cut-off with
$E_{cut}=60 $MeV\cite{Matsuo06}.
However 
this parameter set leads to neutron pairing gaps in Sn isotopes
which are systematically smaller than the empirical values estimated from the odd-even mass difference
using the three-point formula\cite{SDN98}. We therefore vary $\eta$ and fix to 
$\eta=0.71$ so as to reproduce the
empirical gap in $^{120}$Sn\cite{Matsuo07} while keeping the scattering length
relevant to the low-density limit.
We call this parameter set 
DDDI-bare' since it is refereed to the bare nucleon force, but with a small
modification.  

2) The second is the simplest density-independent delta interaction:
\begin{equation}
V_n(\vecr)=v_0. 
\end{equation} 
The only relevant parameter $v_0$ is fixed to $v_0=-195$ MeVfm$^{3}$
to reproduce
the gap in $^{120}$Sn, as is done in Ref.\cite{DD-Dob,DD-mix2}. The density-independent delta interaction
is often called the volume pairing interaction\cite{DobHFB2}. 

3) The third is the so-called mix pairing interaction\cite{DD-mix1,DD-mix2}, which
has a linear dependence on the total nucleon density $\rho(r)$, and given by 
\begin{equation}
V_n(\vecr)=v_0 \left(1-\frac{1}{2}\left(\frac{\rho(\vecr)}{\rho_0}\right)\right)
\end{equation} 
with $\rho_0=0.16$fm$^{-3}$. Here the parameter $v_0$ is fixed to reproduce the
gap in $^{120}$Sn, leading to $v_0=-292$ MeVfm$^{3}$.

The pairing force strength $V_n(\vecr)$ 
is compared among the three kinds of interaction in Fig.\ref{fig:DDDI}, which 
plots $V_n(r)$ as a function of the radial coordinate in
$^{134}$Sn. 
It is seen that the 
DDDI-bare' exhibits the strongest radial dependence, i.e., the strongest
density dependence, and that the interaction strength
in the 
region outer than the nuclear surface 
( corresponding to the low density limit
$\rho(r) \rightarrow 0$ )
is more than twice larger than that of the
volume pairing. 
The mix pairing exhibits an intermediate radial- (density-) dependence.
Since these three versions of DDDI have very different force strength in the
external region we expect to see
the difference in physical observables related to the surface region and/or in the
exterior of nuclei.

\begin{figure*}[htb]
\includegraphics[scale=0.5,angle=270]{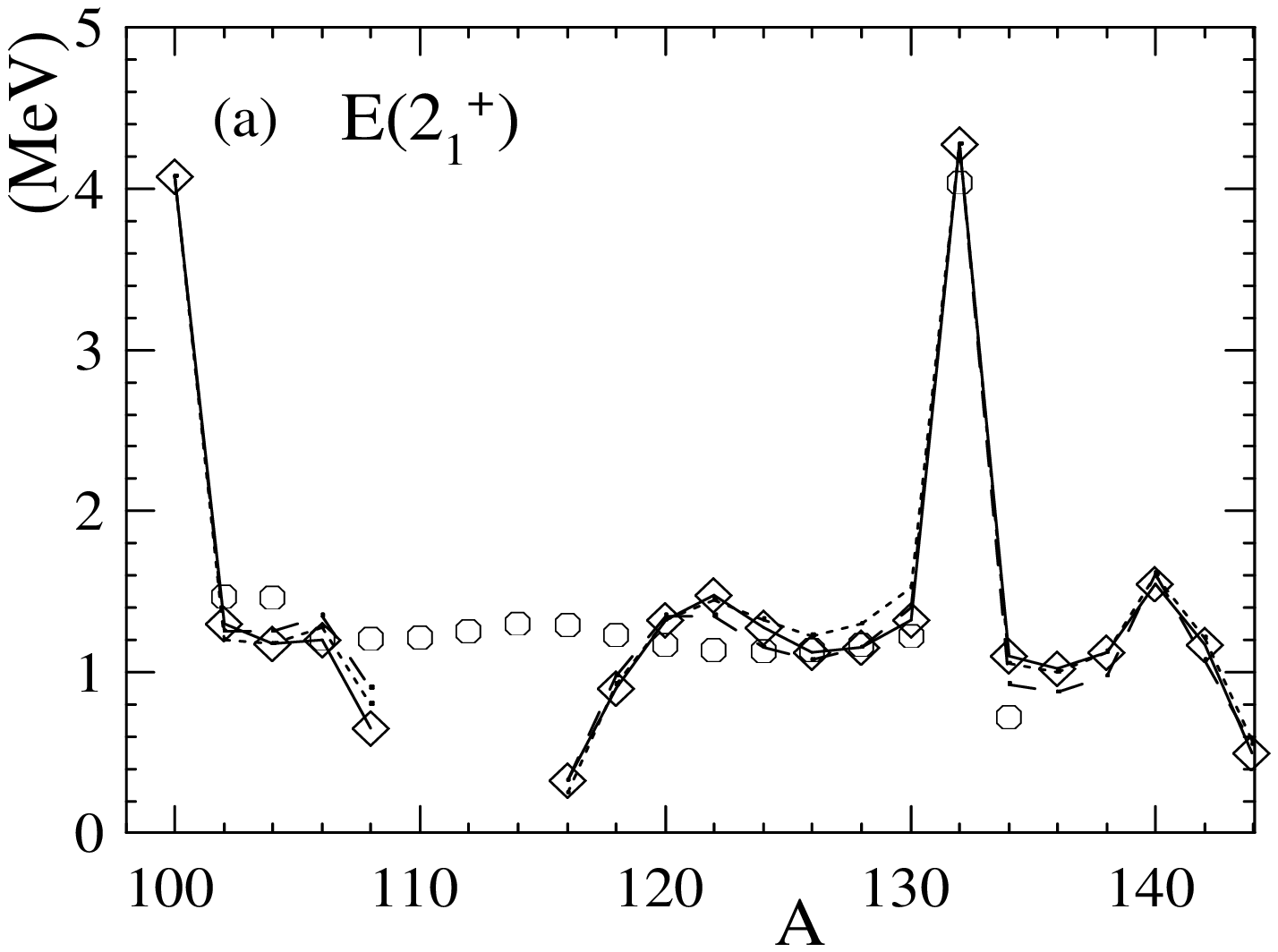}
\includegraphics[scale=0.5,angle=270]{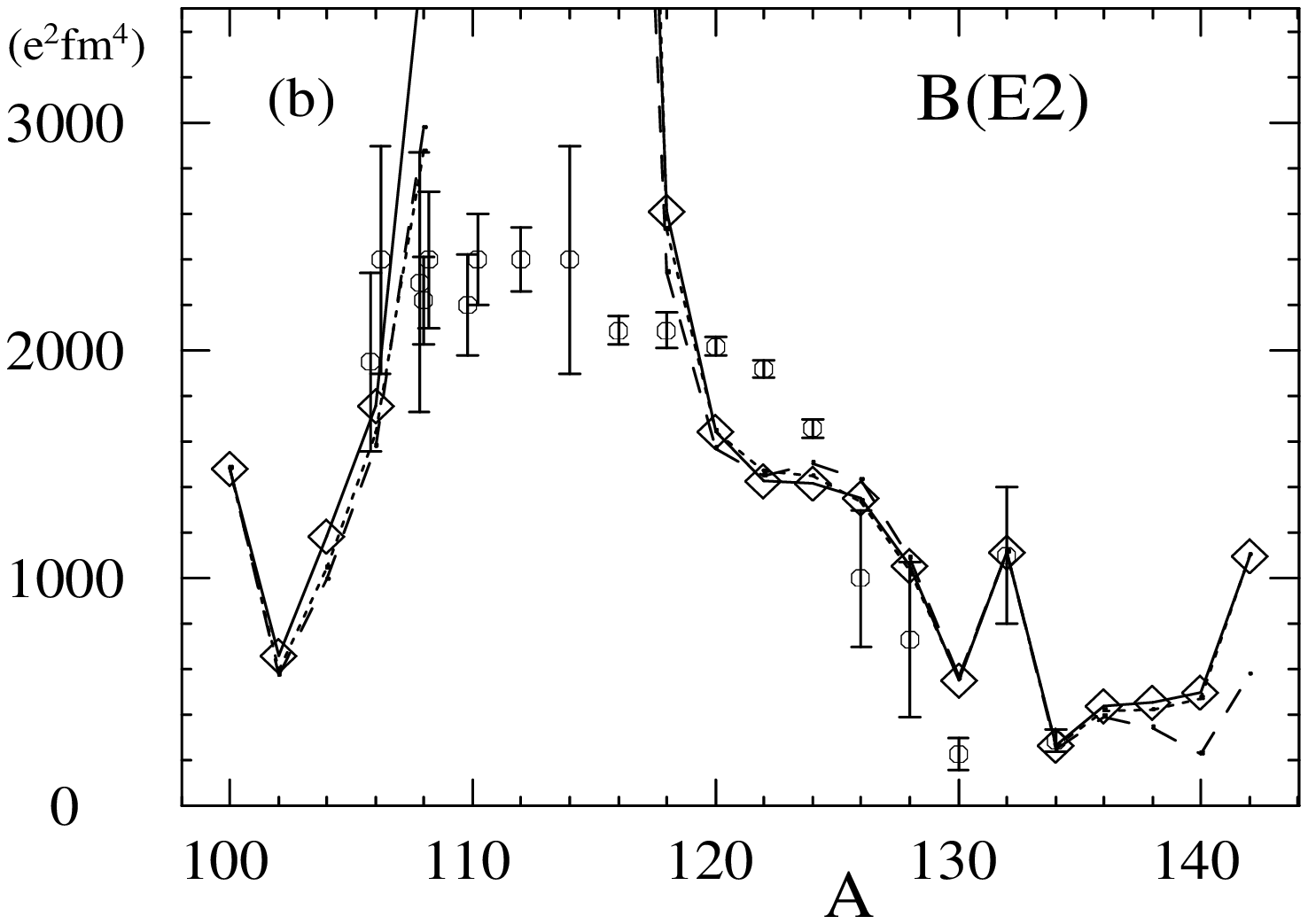}
\includegraphics[scale=0.5,angle=270]{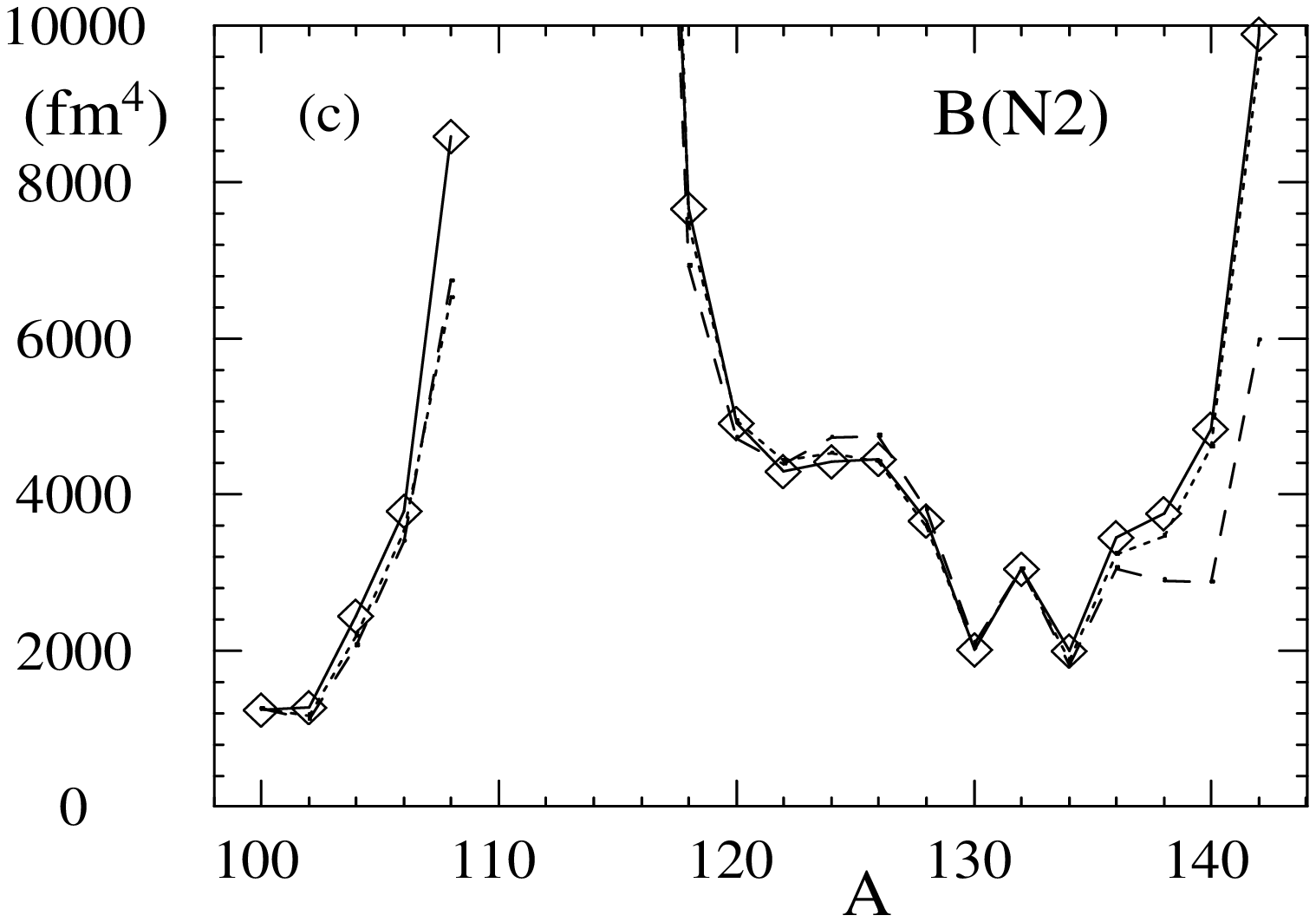}
\includegraphics[scale=0.5,angle=270]{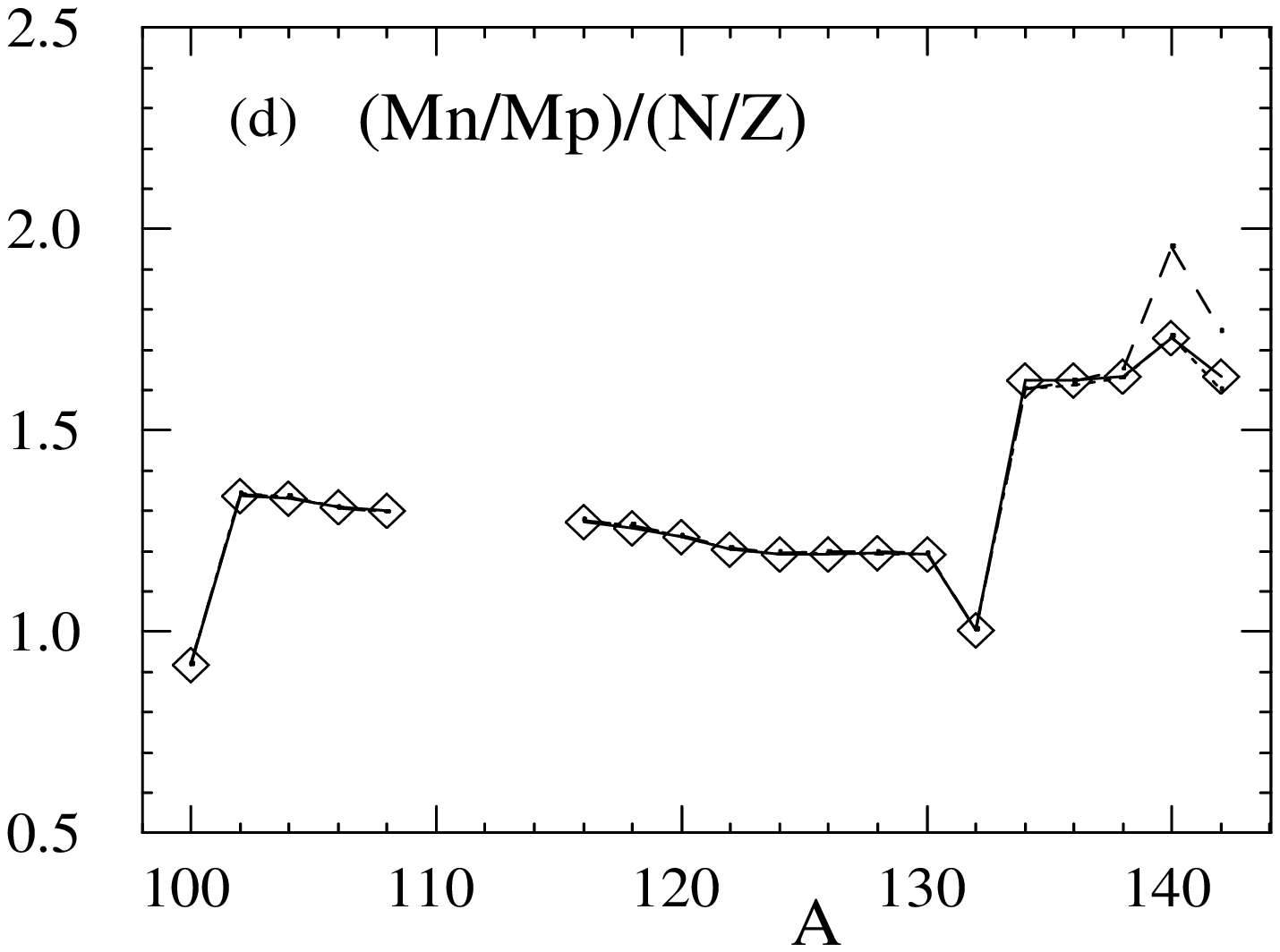}
\caption{\label{fig:Exstrength} %\label{fig:ExBE2} 
(a) The excitation energy of the calculated $2_1^+$ QRPA solutions obtained with
 the DDDI-bare', the mix and the volume pairing interactions, plotted
with the diamonds connected by the solid line, the dotted line, and the
dashed line, respectively. The experimental $2_1^+$ energy\cite{Raman-BE2} is plotted with
the circles. (b) The same but for 
$B({\rm E2}; 0_{{\rm gs}}^+ \rightarrow 2_1^+)$.
The experimental data is taken from Refs.\cite{Raman-BE2,Radford-SnBE2,
Banu-SnBE2,Vaman-SnBE2,Cederkall-SnBE2,Ekstrom-SnBE2}
(c) The same but for the neutron quadrupole strength 
$B({\rm N2}; 0_{{\rm gs}}^+ \rightarrow 2_1^+)$, and (d) for
the ratio $(M_n/M_p)/(N/Z)$ of the neutron and proton amplitudes
associated with the $2^+_1$ QRPA solution.
}
\end{figure*}

We compare in Fig.\ref{fig:gap} the calculated average neutron pairing gap
$\Delta_{n}\equiv 
\int \Delta(\vecr)_n\rhot_n(\vecr)d\vecr/\int \rhot_n(\vecr)d\vecr$
obtained with the three interactions. 
It is seen that they give the average pairing gaps which are almost identical 
for $A \sim 116-130$.
In neutron-rich nuclei $A=134-138$
beyond the $N=82$ shell closure, a
small difference of the order of $<150$ keV is seen 
(except the subshell nucleus $^{140}$Sn where the volume
pairing gives the vanishing gap). It is pointed out 
in the literature\cite{DD-Dob,DD-mix1,DD-mix2,Yam05}
that the surface enhancement of the pairing 
interaction influences the average paring gap only to a small
extent unless one sees the trend in a very wide range of
mass\cite{DD-mix1,DD-mix2}, and the difference may grow only in very 
neutron-rich nuclei close
to the drip-line\cite{DD-Dob,Yam05}. On the other hand, it is 
seen that the DDDI-bare'
gives an overestimate of the average pairing gap for the isotopes
with $A \lesim 116$, by about $\gesim 200$ keV compared with the
other two interactions. For a fair comparison in the following analysis, 
we modify slightly
the value of $\eta$, i.e.,
$\eta=0.75$, only for $A<116$ so that it gives the average pairing gap 
comparable with those of the volume  and the mixed pairing (cf. the
thin solid line in Fig.\ref{fig:gap}).

\section{Results}

In the first two subsections we will mainly discuss results 
which we obtain with the effective pairing interaction DDDI-bare' having
the strong increase of the interaction strength in the surface and the
external regions.
A comparative analysis using 
different pairing interactions will be presented in the  
subsection~\ref{sec:sensitivity}.

\subsection{The lowest-lying quadrupole state}

Let us first discuss the basic properties of the calculated
lowest-lying quadrupole state, on which we focus in the present paper.
Figure \ref{fig:Exstrength}(a) and (b) shows 
the excitation energy $E(2_1^+)$ and $B({\rm E}2; 0_{{\rm gs}}^+ \rightarrow 2_1^+)$ 
of the $2_1^+$ QRPA solution in the Sn isotopes. 
The results 
are similar to those in 
the
full self-consistent QRPA calculation \cite{terasaki06} using the same
Skyrme interaction SLy4. 
It is seen in Fig.~\ref{fig:Exstrength}(a) and (b) that 
the calculated $E(2_1^+)$ and $B(E2)$ reproduce reasonably well the
experimental values. In the present QRPA
calculation the lowest quadrupole mode shows an instability 
in the $A=110-114$ and $A>144$ isotopes as is also 
seen in Ref.\cite{terasaki06}. We do not discuss these cases 
and the $A=116, 144$ cases where the QRPA excitation energy is
less than 0.5 MeV since the quadrupole mode then needs to be
described as an anharmonic vibration, which is beyond the scope of the
present work. Note that there exist a few other QRPA calculations 
for the neutron-rich Sn\cite{Ansari06,Severyukhin08}.

\begin{figure}[tb]
\includegraphics[scale=0.4,angle=270]{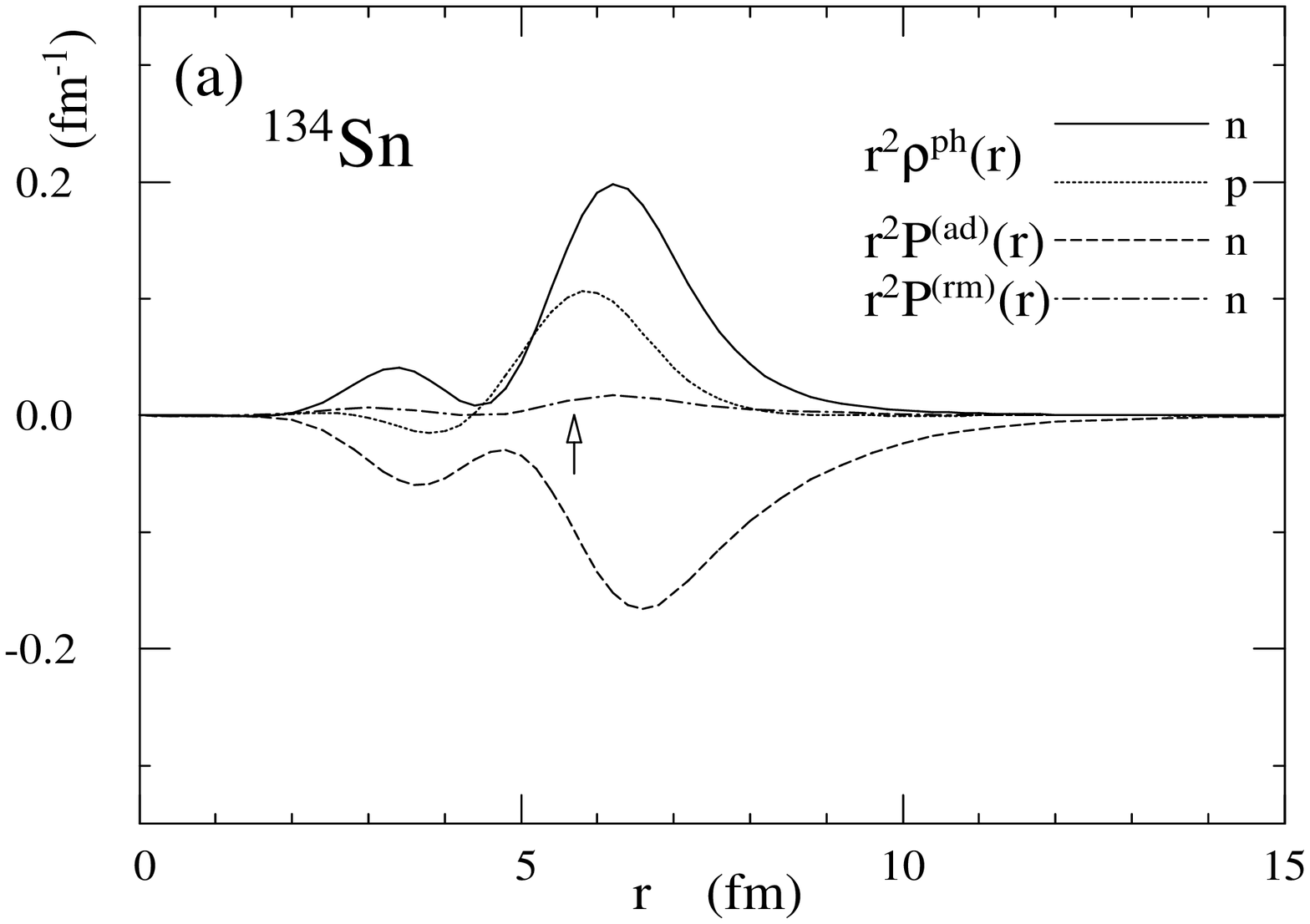}
\includegraphics[scale=0.4,angle=270]{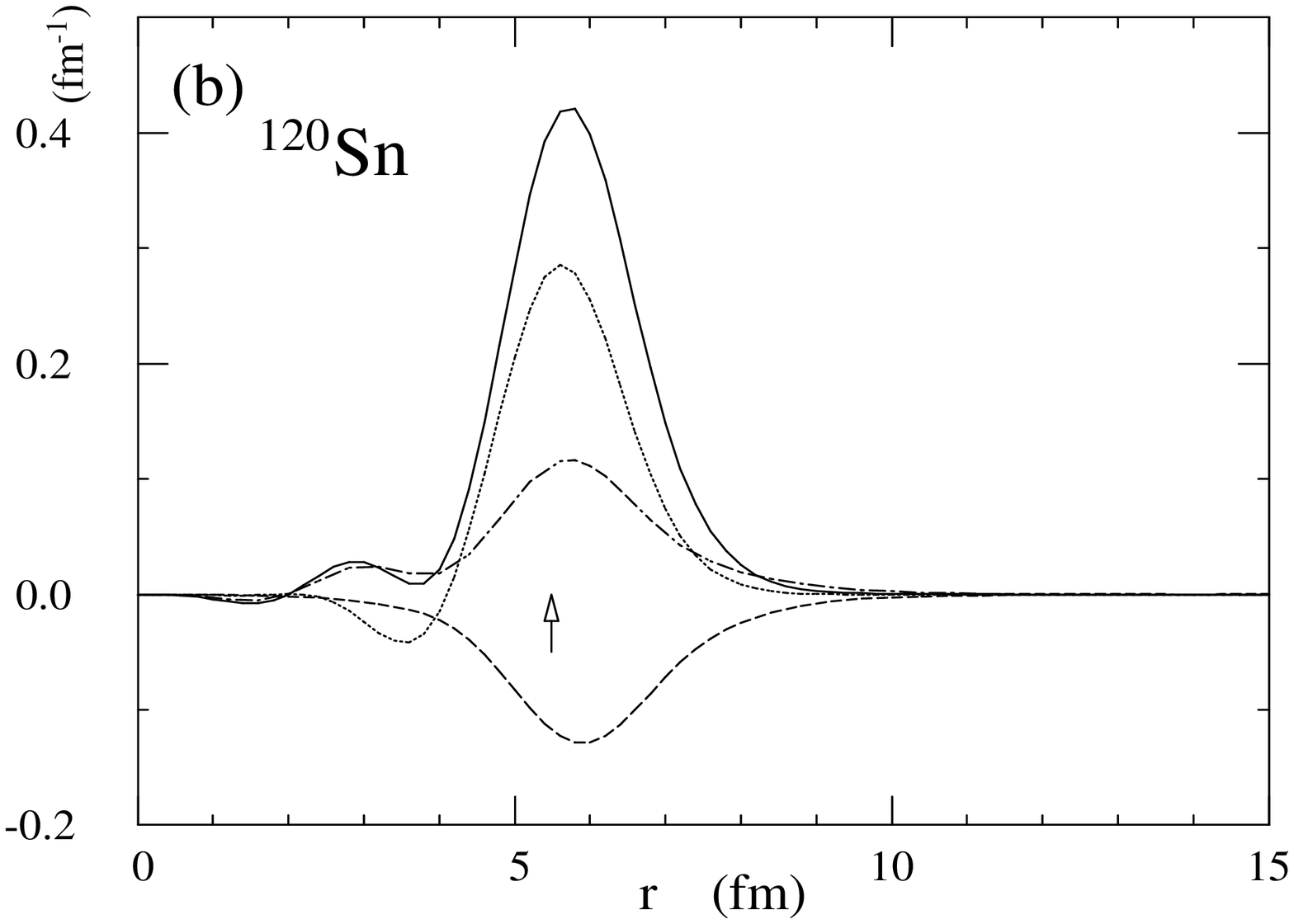}
\caption{\label{fig:trans-sn120-sn134} 
The transition densities associated with the $2_1^+$ QRPA solution (a) in
$^{134}$Sn and (b) in $^{120}$Sn. The solid and the dotted curves are the particle-hole
transition density
$r^2\rho^{ph}(r)$ for neutrons and protons respectively while
the dashed and the dotdashed curves represent  the pair-addition transition
density
$r^2P^{({\rm ad})}(r)$ and the pair-removal transition density 
$r^2P^{({\rm rm})}(r)$ of neutrons, respectively. The arrow indicates 
the half density radius $R_{1/2}$. The effective pairing interaction is the
DDDI-bare'.
}
\end{figure}

The lowest-lying quadrupole state in isotopes beyond the $N=82$ shell
closure are of special interest. Figure \ref{fig:Exstrength}(c) shows the
neutron quadrupole strength
$B({\rm N}2; 0_{{\rm gs}}^+ \rightarrow 2_1^+)$  calculated in the same way as 
$B({\rm E}2)$, but for neutrons
instead of protons. The isotopic dependences
of $B({\rm E}2)$ and $B({\rm N}2)$ look similar for $N<82$, 
but the values of $B({\rm N}2)$ are apparently larger than the 
proton strength $B({\rm E}2)/e^2$ as 
the neutron number exceeds the magic number $N=82$.
This behavior is seen clearly in the ratio 
$M_n/M_p\equiv \sqrt{B({\rm N}2)/(B({\rm E}2)/e^2)}$
of the neutron and proton amplitudes, plotted in 
Fig.~\ref{fig:Exstrength}(d).  If the low-lying quadrupole mode
is a collective surface oscillation of liquid drop of 
uniform mixture of neutrons and protons, 
the ratio $M_n/M_p$ would be $N/Z$. The calculated double ratio
$(M_n/M_p)/(N/Z)$ is around 1.2-1.3 for $A<132$, being not very different
from the liquid drop limit 1, and it is consistent with the
typical values ($\approx 1.3-1.5$) evaluated experimentally 
in stable isotopes $^{116-124}$Sn\cite{Kennedy92}.
The double ratio, however, increases suddenly in $A>132$, and 
amounts to 1.6-1.7.
The ratio $M_n/M_p$ itself reaches 2.7-3.1. In the isotopes beyond $N=82$,
the amplitude $M_n$ of neutrons dominate over 
that $M_p$ of protons in the first $2^+$ state. 
Note that the $(M_n/M_p)/(N/Z)$ ratio obtained in 
Ref.\cite{Severyukhin08,Severyukhin04} is somewhat smaller than our
values by about 0.2.

A characteristic feature of the lowest-lying quadrupole states beyond the 
$N=82$ shell gap can be demonstrated more explicitly in terms
of 
the transition densities, which are shown in Fig.\ref{fig:trans-sn120-sn134} 
for (a) $^{134}$Sn and (b) $^{120}$Sn. 
In $^{120}$Sn, the transition densities
exhibit the features typical of the low-lying quadrupole surface
vibration. Namely, the  particle-hole transition densities for neutrons and protons,
and the pair-addition 
and -removal transition densities of neutrons exhibit similar radial
profile all of which have peaks 
around the nuclear surface
$r\sim R_{1/2}\approx 5.5$ fm ($R_{1/2}$ denotes the half-density
radius defined by $\rho(r)=0.08$ fm$^{-3}$). 
In $^{134}$Sn, in contrast, the peak positions
of the
 neutron and proton particle-hole transition density
 $\rho^{ph}_{n,p}(r)$ 
deviates
significantly with each other:
the neutron transition density has amplitude which 
extends toward the exterior of the nucleus while the proton transition
density does not.
%(This is a sort of decoupling between neutrons and protons,
%and it can be related to the large excess in $M_n/M_z$
%with respect to the liquid drop value $N/Z$. )

The transition density $P^{({\rm ad})}(r)$ of the 
neutron pair-addition mode 
is most characteristic in $^{134}$Sn.
It has the largest amplitude 
in the external region 
$r\gesim R_{1/2}$ ($\approx 5.7$ fm) 
among the three kinds of transition 
densities.
The position  where the amplitude of $P^{({\rm ad})}(r)$
is maximum, $r\approx 6.6$ fm, deviates by about 1fm 
from the nuclear surface ($r \approx 5.7$ fm),  and it is even
outside of the peak position of the neutron 
particle-hole transition density.
In addition, it is extended significantly toward the outside of the nucleus,
and the tail reaches far to 12-13 fm (in the  scale of the figure).
The extension of the amplitude in the external region 
is much more significant than that 
in the neutron particle-hole transition density $\rho^{ph}(r)$.
We note also that the transition density $P^{({\rm rm})}(r)$ of 
the neutron pair-removal mode is very small in comparison with 
the pair-addition transition
density $P^{({\rm ad})}(r)$, and the dominance of the pair-addition
amplitude is commonly seen in all the
cases of $^{134-142}$Sn.

\begin{figure*}[htb]
\includegraphics[scale=0.4,angle=270]{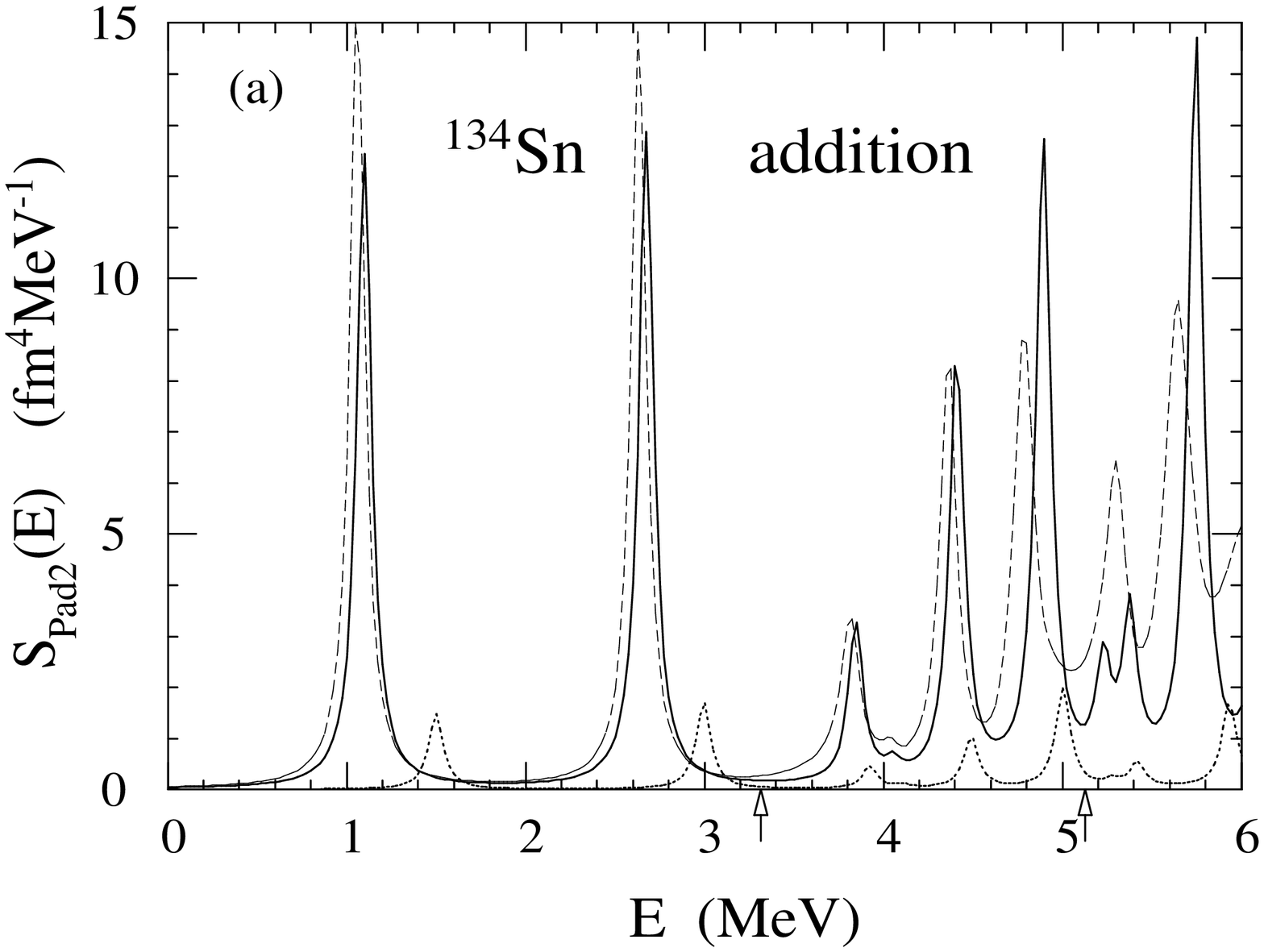}
\includegraphics[scale=0.4,angle=270]{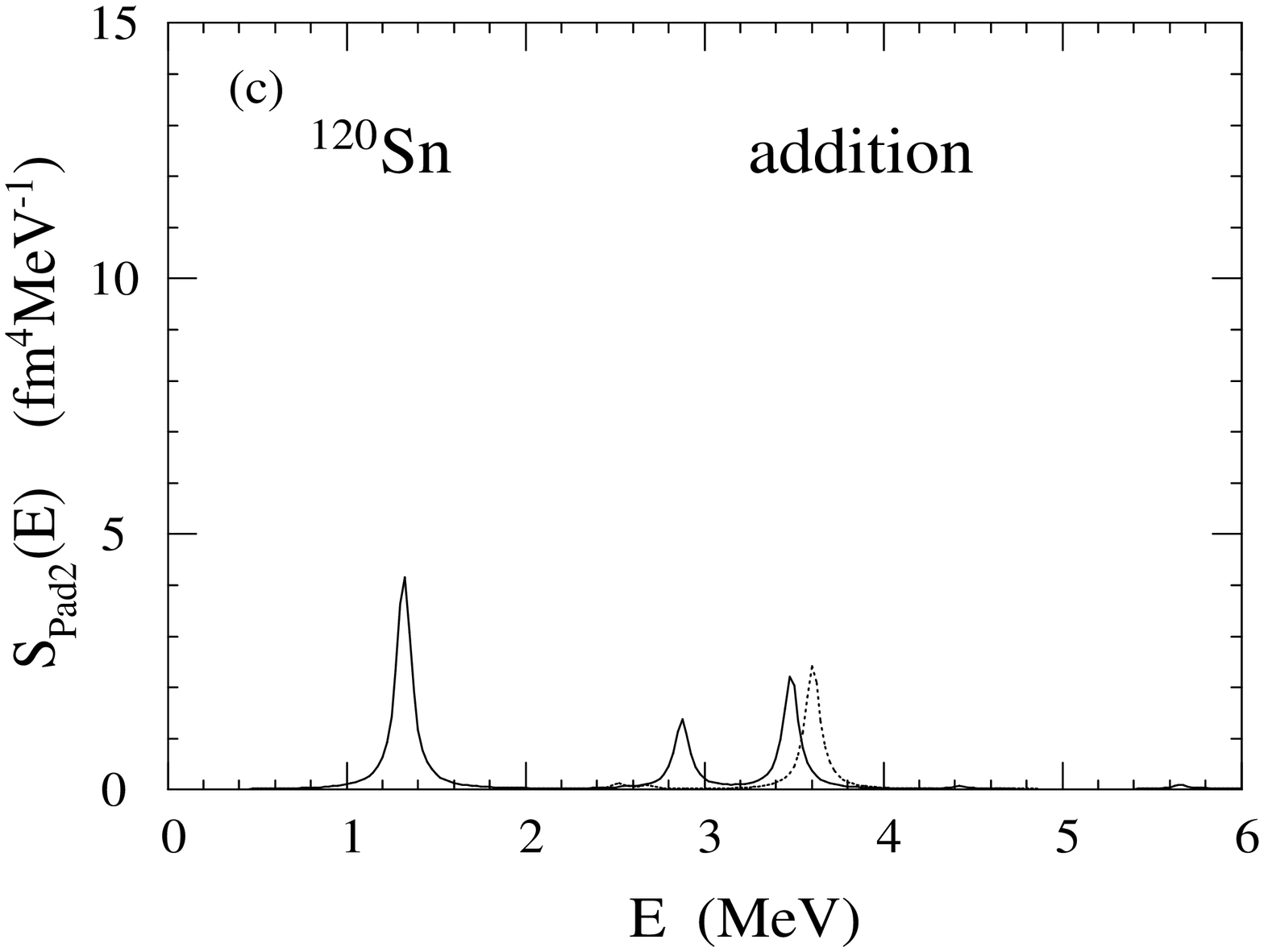}
\includegraphics[scale=0.4,angle=270]{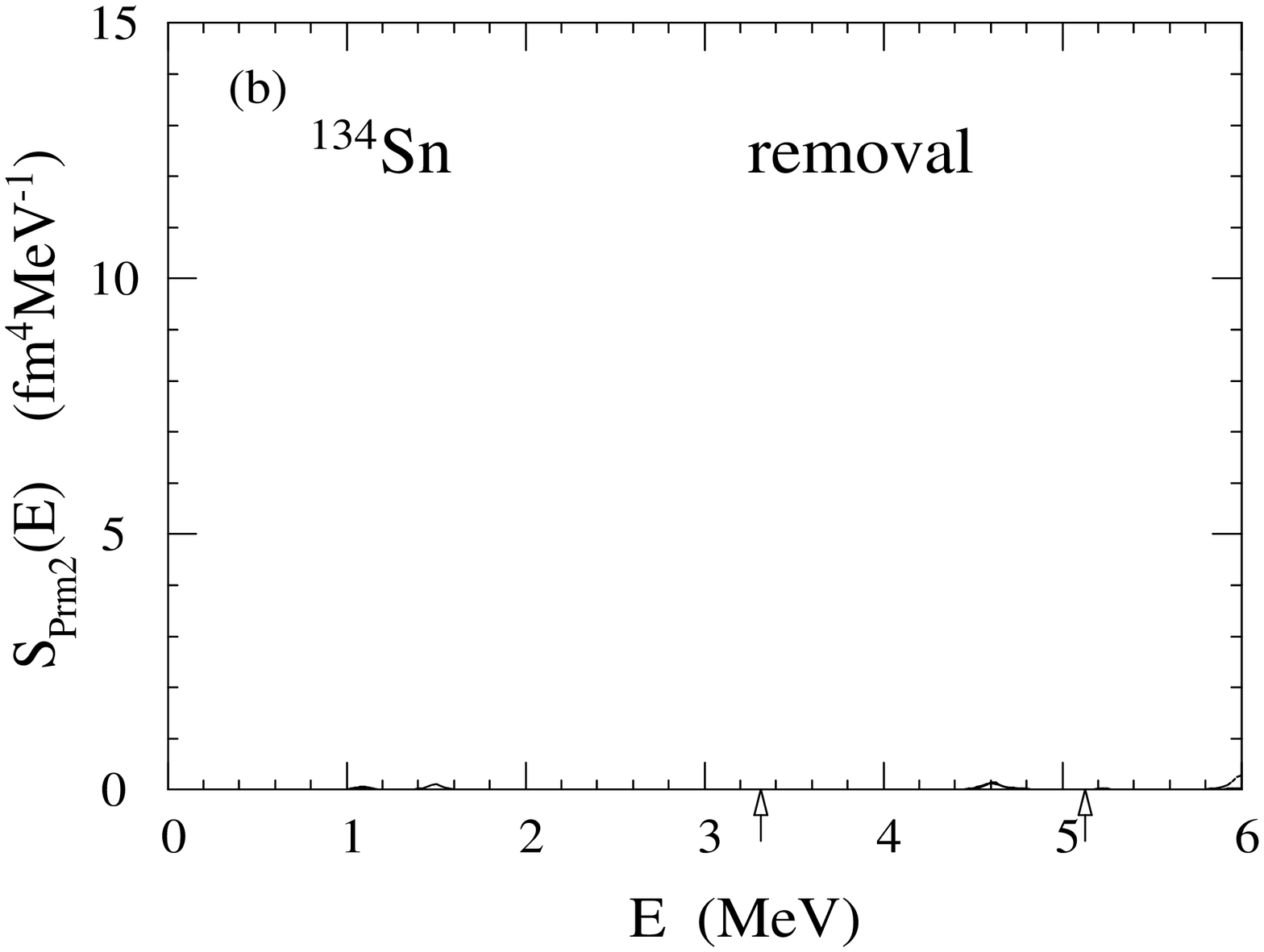}
\includegraphics[scale=0.4,angle=270]{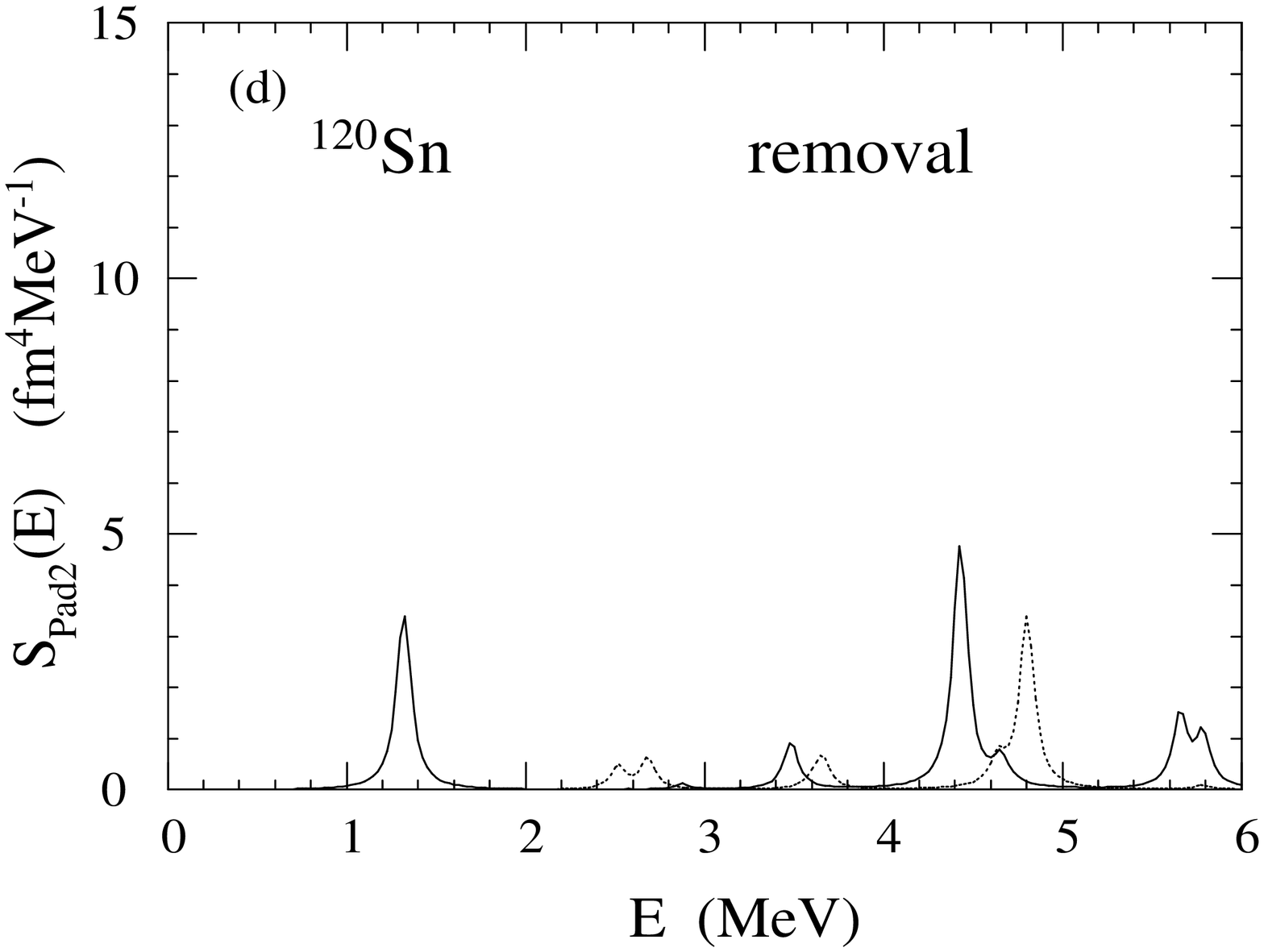}
\caption{\label{fig:pstr-sn120-sn134} 
(a) The transition strength function $S_{{\rm Pad}2}(E)$  for the 
quadrupole pair-addition mode in $^{134}$Sn,
plotted with the solid curve. The thin dashed curve is the result of the
continuum QRPA calculation while the dotted curve represents the 
unperturbed response for the two-quasiparticle excitations. 
The effective pairing interaction is the DDDI-bare'. 
The smearing parameter is $\epsilon=50$ keV.
The arrows indicate the calculated one- and two-neutron separation energies $S_{1n}$ and
$S_{2n}$.
(b) The same
as (a) but for the transition strength function $S_{{\rm Prm}2}(E)$  for the 
quadrupole pair-removal mode. (c)(d) The same as (a) and (b), but for
$^{120}$Sn.
}
\end{figure*}

\subsection{Quadrupole pair transfer}

\begin{figure}[htb]
\includegraphics[scale=0.5,angle=270]{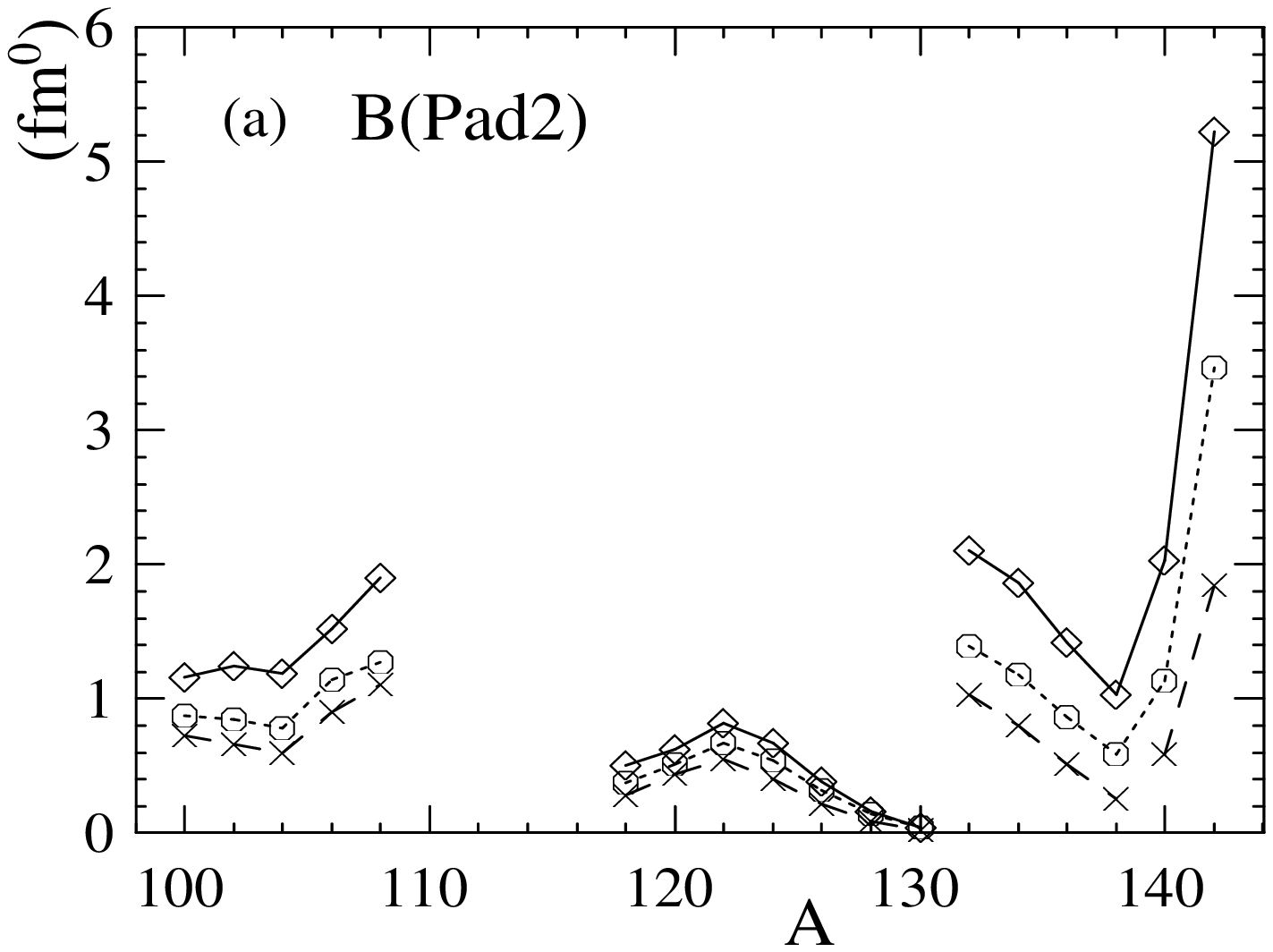}
\includegraphics[scale=0.5,angle=270]{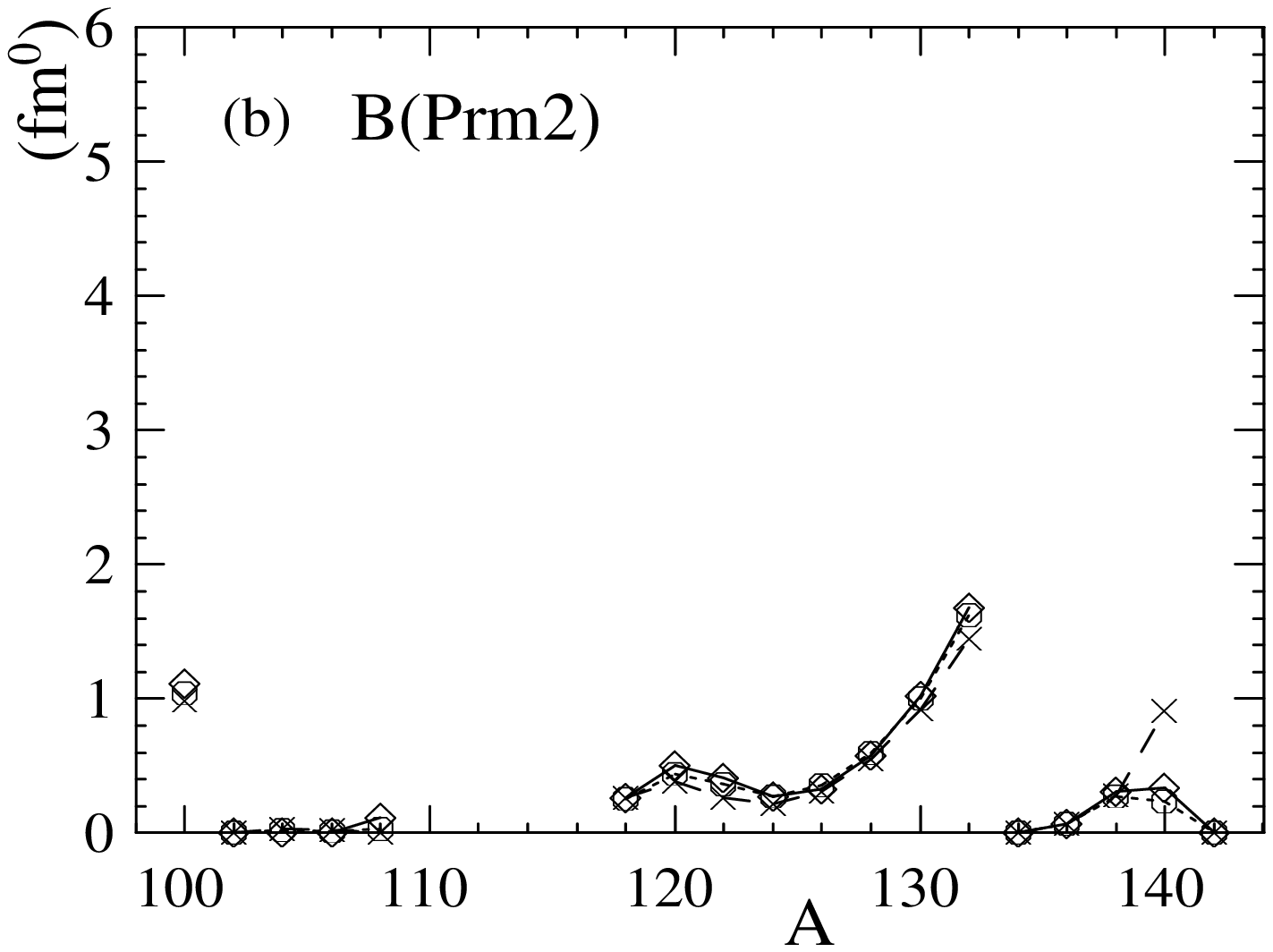}
\caption{\label{fig:pairstrength} 
(a) The calculated transition strength $B({\rm Pad}2)$ of the neutron
pair-addition transfer feeding the $2_1^+$ states in Sn isotopes. 
The diamonds connected with the solid line are the results obtained
 with DDDI-bare' while the dotted and the dashed lines (the circles and
the crosses) are for
the mix and the volume pairing interactions, respectively.
(b) The same but for the strength $B({\rm Prm}2)$ of 
the quadrupole pair removal transfer.
}
\end{figure}
 
\begin{figure*}[htb]
\includegraphics[scale=0.4,angle=270]{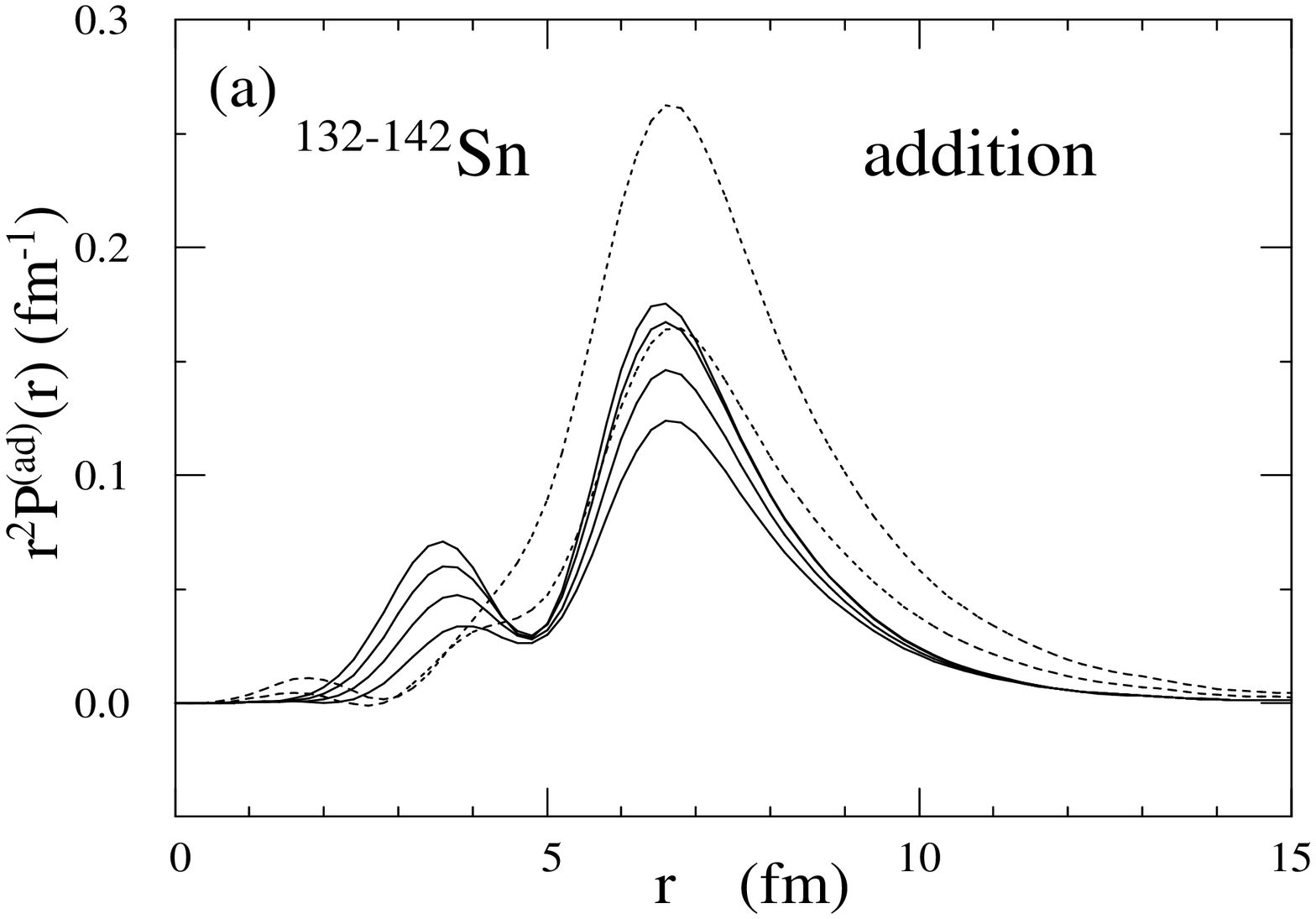}
\includegraphics[scale=0.4,angle=270]{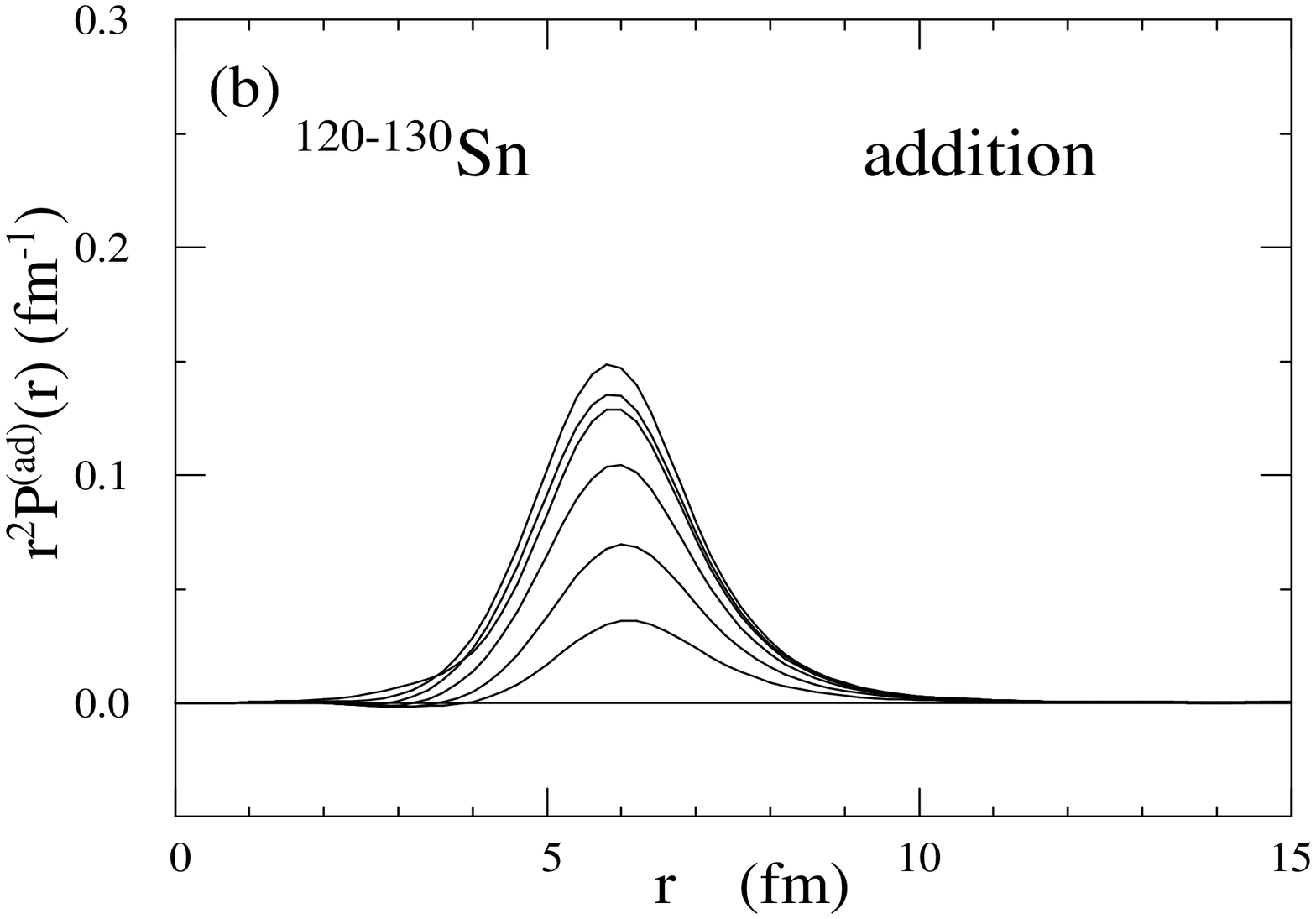}
\includegraphics[scale=0.4,angle=270]{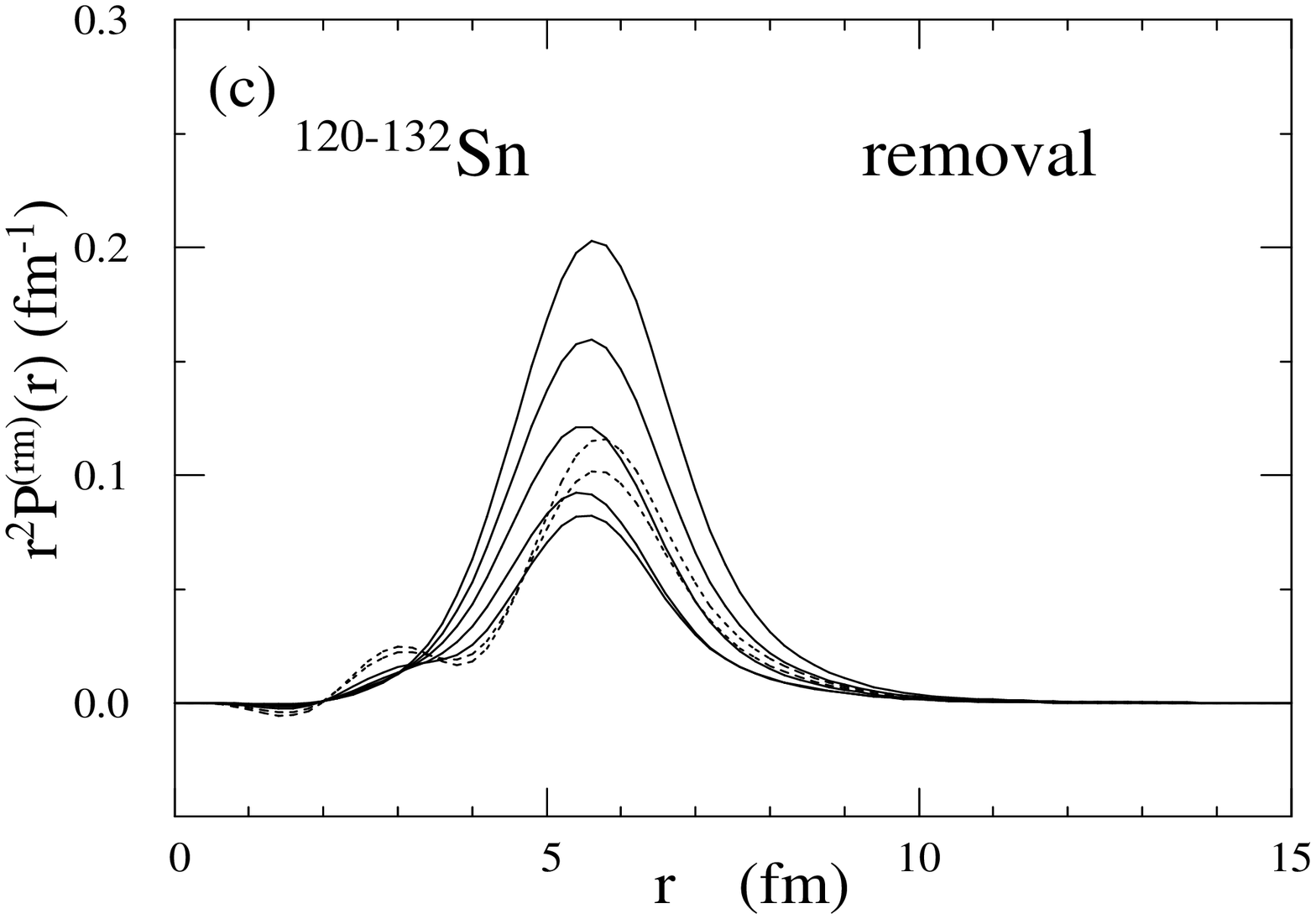}
\caption{\label{fig:transitiondensity} 
(a) The transition density $r^2P^{({\rm ad})}(r)$ for the neutron
 pair-addition transfer associated with
the $2_1^+$ QRPA solutions, calculated in the isotopes $^{132-142}$Sn.
The pair-addition
transition strength $B({\rm Pad}2)$ is used to normalize, and
the phase is chosen so that
the transition matrix element becomes positive.
 The solid curve is for  $^{120-132}$Sn while the dotted 
curve is used for $^{140-142}$Sn. The effective pairing interaction is the DDDI-bare'.
(b) The same as (a), but for $^{120-130}$Sn. 
(c) The transition density $r^2P^{({\rm rm})}(r)$ for the
neutron  pair-removal transfer
associated with
the $2_1^+$ QRPA solutions, 
calculated  in $^{120-132}$Sn. The solid and the dotted curves
are for 
$^{124-132}$Sn and $^{120,122}$Sn, respectively. 
}
\end{figure*}

Figure \ref{fig:pstr-sn120-sn134} shows the strength 
functions $S_{{\rm Pad}2}(E)$ and  $S_{{\rm Prm}2}(E)$
for the quadrupole pair-addition and -removal modes in $^{134}$Sn and 
$^{120}$Sn. 
The pair-addition strength associated with the
first $2^+$ mode in $^{134}$Sn is significantly larger than 
that in $^{120}$Sn, 
and it reflects the large pair-addition amplitude seen in 
Fig.\ref{fig:trans-sn120-sn134}. 
The unperturbed strength functions are also plotted for 
comparison (the dotted curve).
The first peak  in the unperturbed calculation of $^{134}$Sn corresponds to 
the transition to the two-quasi-neutron 
configuration $[1f_{7/2}]^2_{I=2}$.
The pair-addition strength for the $2_1^+$ state
in the full calculation is about ten times larger than that in
the unperturbed calculation, thus it indicates a significant enhancement
due to a collectivity (the RPA correlation in the other words) 
in the pair-addition mode. 
Another feature observed in $^{134}$Sn is that the RPA correlation effect is large 
not only for the lowest mode  but also for  higher lying
modes: 
the second peak  of $S_{{\rm Pad}2}(E)$ also has a large pair-addition
strength comparable with the first peak, and there exist several peaks in the excitation energy region
$E \gesim 4$ MeV 
with significant pair-addition strength. The peaks 
above the one-neutron separation energy $S_{1n}$ are discretized continuum
states. The continuum QRPA calculation\cite{Matsuo01,Serizawa09} is appropriate
for the excitation energies above $S_{1n}$. In fact, the result of the continuum
QRPA calculation, shown with the thin dashed curve in Fig.~\ref{fig:pstr-sn120-sn134},
indicates that the proper treatment of the continuum modifies the strength function
for $E> S_{1n}$. On the other hand, it has only a minor effect on the
strengths below $S_{1n}$, and it does not affect the discussions below. 
Note also that the pair-removal strength in $^{134}$Sn is extremely small
in the displayed energy range.

\subsubsection{pair-addition mode}

We shall discuss  properties of the pair-addition mode in detail.
Figure \ref{fig:pairstrength}(a) is the systematics of the pair-addition strength $B({\rm Pad}2)$ 
associated with the $2_1^+$ mode.  
A feature immediately noticed in the figure 
is a strong variation along the isotopic chain.
 $B({\rm Pad}2)$ takes a rather small value for 
$A<132$, but it
shows a big jump at $A=132$, where the strength is larger
than those in any other isotopes with $A<132$.  
Proceeding from $^{132}$Sn to$^{134}$Sn, $^{136}$Sn and $^{138}$Sn,
the strength decreases gradually while keeping
relatively large values. It then
turns to increase  in $^{140}$Sn. and takes a 
very large value
in $^{142}$Sn.

Figure~\ref{fig:transitiondensity}(a) shows 
the pair-addition transition densities $P^{({\rm ad})}(r)$ associated
with the lowest quadrupole mode in the isotopes $^{132-142}$Sn beyond the 
$N=82$ shell closure. It is seen 
that they all exhibit sizable amplitudes  extending to $r\gesim 8$ fm 
as is found in Fig.\ref{fig:trans-sn120-sn134} in the case of $^{134}$Sn.
The spatial extension of the pair-addition transition density 
develops further in $^{140}$Sn and $^{142}$Sn, for which we see 
significant amplitude
even around $r \sim 12-13$ fm, which is about $\sim 7$ fm apart from 
the half density surface $R_{1/2}= 5.6-5.8$ fm.
This is contrasted with 
the pair-addition transition density for
the isotopes $^{120-130}$Sn below the shell closure (displayed  in 
Fig.\ref{fig:transitiondensity}(b)).  In these nuclei
spatial extension is hardly seen in the pair-addition transition
density, and the transition densities is mostly concentrated in the
internal and surface region $r \lesim R_{1/2} +2$ fm $\approx 8$ fm.  
The spatial extension
emerges suddenly at $A=132$.  
Comparing the pair-addition transition density in $^{122}$Sn 
(the largest one in Fig.\ref{fig:transitiondensity}(b))
with that in $^{136}$Sn (the third largest one among the solid curves in
Fig.\ref{fig:transitiondensity}(a)), the maximum amplitude is
almost the same, but 
the pair-addition strength $B({\rm Pad}2)$ in $^{136}$Sn 
($B({\rm Pad}2)=1.42$) is about twice as large as
that in $^{122}$Sn ($B({\rm Pad}2)=0.82$) because of
the spatial extension in $^{136}$Sn.

The sudden change across the $N=82$ shell closure
may be ascribed to 
the fact that the isotopes with $A>132$ have a feature of
weakly bound nuclei although they are located 
far from the neutron drip-line.
The calculated one-neutron separation energy suddenly change
from $S_{1n}=7.62$ MeV in $^{132}$Sn to 
$S_{1n}=3.32$ MeV in $^{134}$Sn.
This arises from the shell gap at $N=82$ and from the fact that 
the neutron single-particle orbits $1f_{7/2}$ and 
$3p_{3/2}$ lying above the
$N=82$ shell gap 
are bound only weakly ( the Hartree-Fock single-particle energies
in  $^{132}$Sn are $e_{1f_{7/2}}=-1.99$ MeV and $e_{3p_{3/2}}=-0.25$ MeV). 
These orbits have 
significantly small binding energies compared with 
the orbit $h_{11/2}$  just below the shell gap ($e_{h_{11/2}}=-7.68$ MeV).
Since the quadrupole pair-addition mode in 
$^{132}$Sn and heavier isotopes 
necessarily contain two-quasiparticle 
components consisting of these weakly bound orbits, the
pair-addition transition density should inherit the long tail arising
from the weak binding.   In particular, 
approaching $^{140}$Sn and $^{142}$Sn, neutrons
start to occupy the $3p_{3/2}$ orbit which has a very small binding energy.

We remark that the closing of shells has a strong influence
on the pair transfer in general\cite{Broglia73}. For instance, 
the neutron shell is closed at $A=132$ 
($N=82$) and the calculated neutron pairing gap vanishes there.
In this case,  the QRPA  equation
(\ref{rpa}) is decomposed into the particle-hole, the particle-particle, 
and the hole-hole sectors.
The pair-addition mode in $^{132}$Sn belongs to the
particle-particle sector, and the lowest-lying QRPA solution 
in this sector has a character of a correlated pair of neutrons built
upon the $^{132}$Sn core.
It is nothing but 
the quadrupole pair vibrational mode in closed shell nuclei, and in this
case a large collectivity in the pair-addition mode is known, 
with a typical example of 
$^{208}$Pb\cite{Bes-Broglia71,Broglia73,Broglia-Bes-Nilsson74}.
This is, however, only one of the mechanisms of the large pair-addition strength
$B({\rm Pad}2)$ seen at $A=132$  ($N=82$).  We note that
the strength in $^{132}$Sn for the DDDI-bare' interaction 
is more than twice larger than that in $^{100}$Sn (another shell
closure at $N=50$), and this 
relative enhancement at $^{132}$Sn is a 
consequence of the characteristic spatial
extension of the transition density in the isotopes beyond
the $N=82$ shell closure.

\subsubsection{pair-removal mode}

Figure~\ref{fig:pairstrength} (b) shows the systematics of the 
pair-removal strength $B({\rm Prm}2)$
for transition from the ground state of an isotope $A$ to 
the $2_1^+$ state in the  isotope $A-2$.
It is seen that the pair-removal strength
is generally small, compared with the pair-addition strength.
%, except for 132Sn. 
Another trend seen here is that the pair-removal strength is 
very small at the beginning of shells
(at $A\approx 102,134$) and of a subshell (at $A=142$); it takes
a value comparable with the pair-addition strength 
$B({\rm Pad}2)$ around the mid-shell ($A\sim 120$),
and it is relatively large at $A=100,132$ because of the
shell closures at $N=50$ and $N=82$ in parallel to the
situation in $^{208}$Pb.\cite{Broglia73} 
%In the case of DDDI-bare' having the strong enhancement of the
%pairing at low densities, the pair-addition strength 
%B(PA2) dominates over the pair-remove strength
%B(PR2) in all the isotopes with $A>132$ since 
%B(PA2) is enhanced significantly by the spatial extension
%of the transition density.

Figure~\ref{fig:transitiondensity}(c) 
displays the pair-removal transition densities $P^{({\rm rm})}(r)$
associated with the $2_1^+$ mode in the 
isotopes from $^{120}$Sn 
(located in the middle of the shell) to $^{132}$Sn 
(at the end of the shell). 
It is seen that the 
transition densities $P^{({\rm rm})}(r)$ of the pair-removal mode
in these isotopes 
does not exhibit the spatial extension toward the outside,
rather it 
is concentrated only around and inside the surface region 
$r \lesim R_{1/2}+2$ fm $\sim 8$ fm. 
This feature originates from the fact that the pair-removal transition
consists  mainly of the obits situated near the Fermi energy
and those bound more deeply.  In the isotopes with
$A\leq 132$, these are strongly bound orbits 
with the binding energy $e_{s.p.} \lesim -7$ MeV (e.g, 
$e_{h_{11/2}}=-7.68$ MeV in
$^{132}$Sn)
and they do not have tails extending far outside the nuclear surface.

\begin{figure}[htb]
\includegraphics[scale=0.4,angle=270]{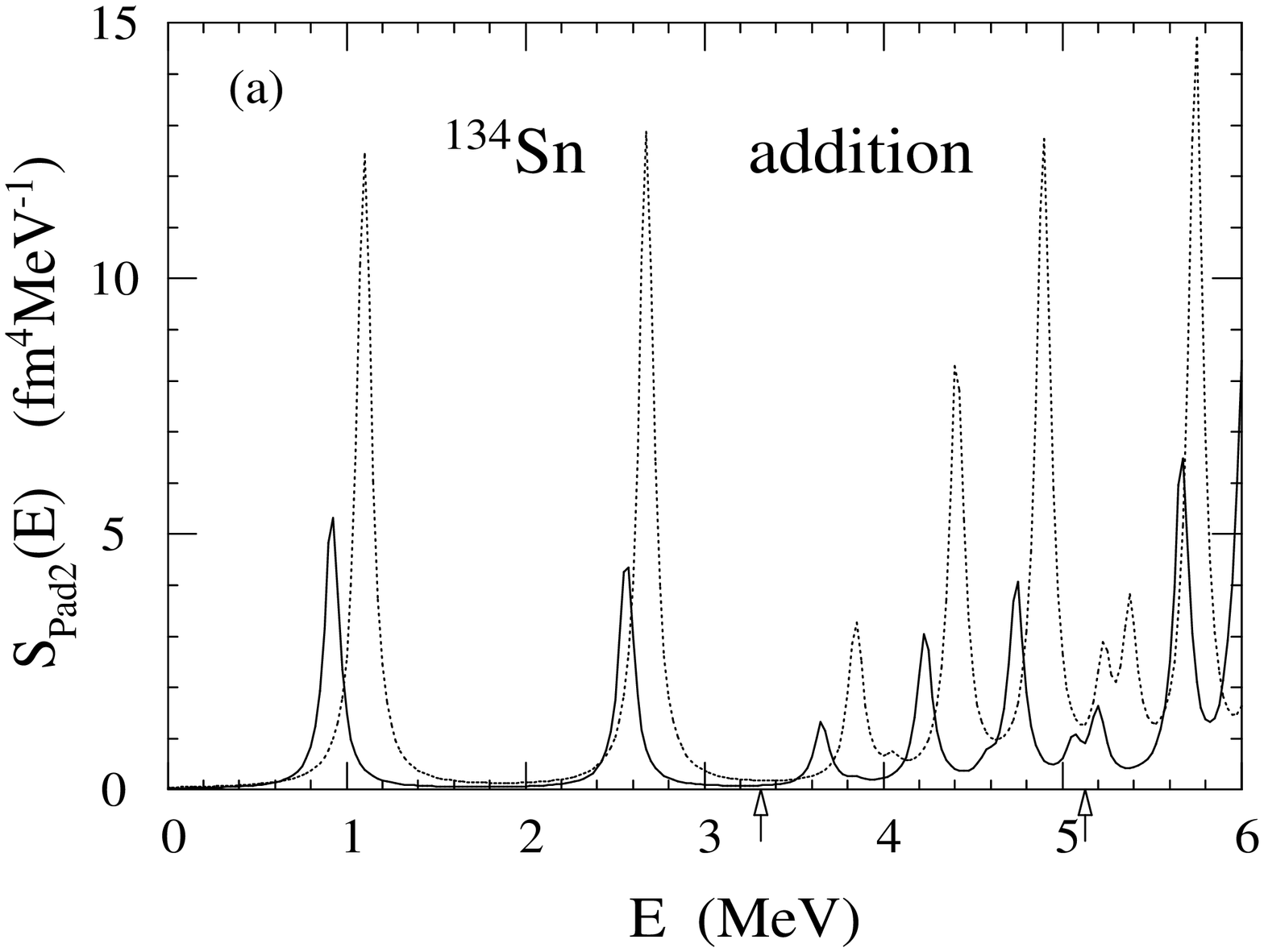}
\includegraphics[scale=0.4,angle=270]{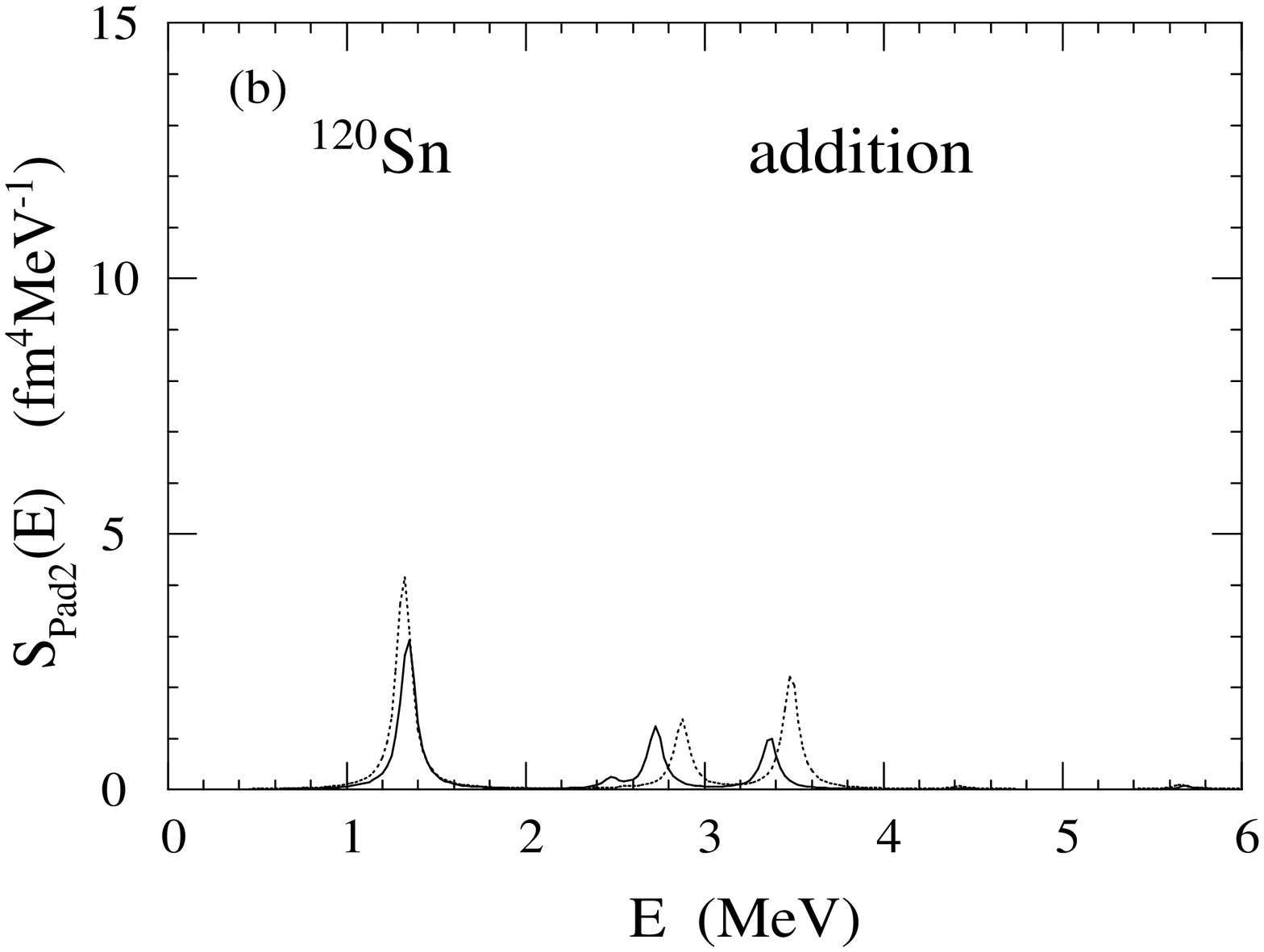}
\caption{\label{fig:pdd-sn120-sn134} 
Comparison of the transition strength functions $S_{{\rm Pad}2}(E)$  for the 
neutron quadrupole pair-addition transfer, calculated with 
the volume pairing interaction
(solid curve) and with the DDDI-bare' (dotted curve). (a) and (b) are for  $^{134}$Sn
and $^{120}$Sn, respectively.
}
\end{figure}

\begin{figure*}[htb]
\includegraphics[scale=0.4,angle=270]{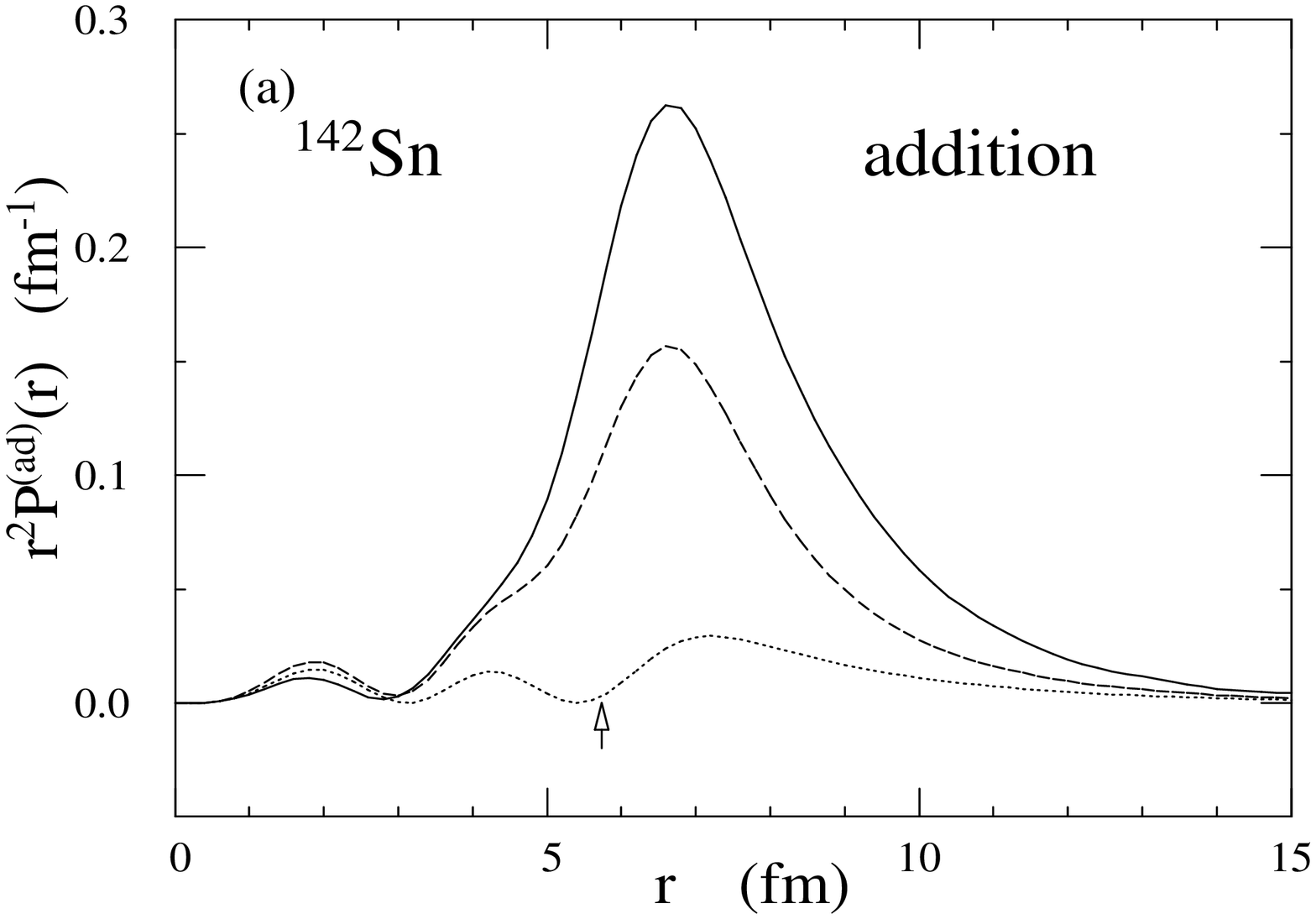}
\includegraphics[scale=0.4,angle=270]{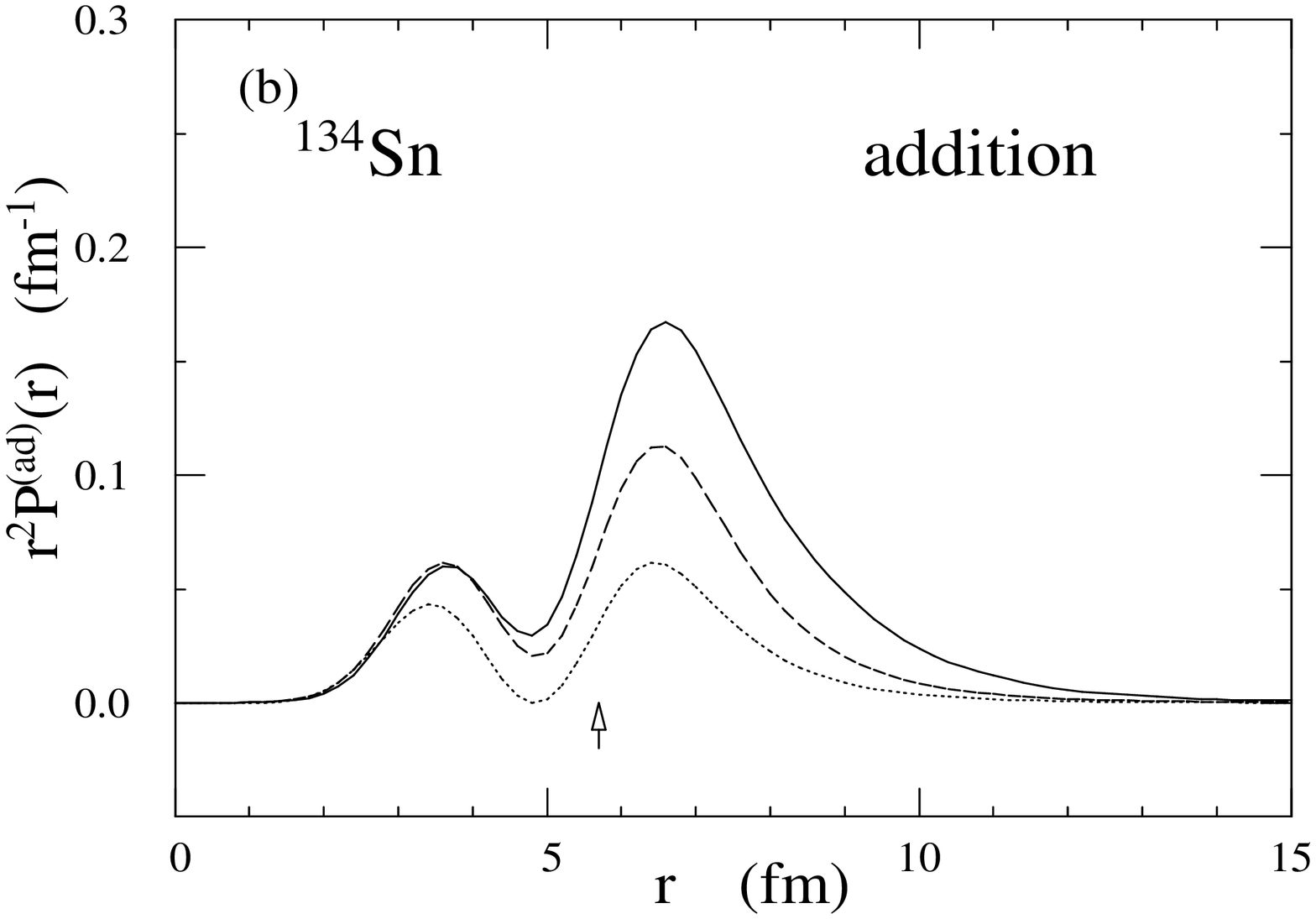}
\includegraphics[scale=0.4,angle=270]{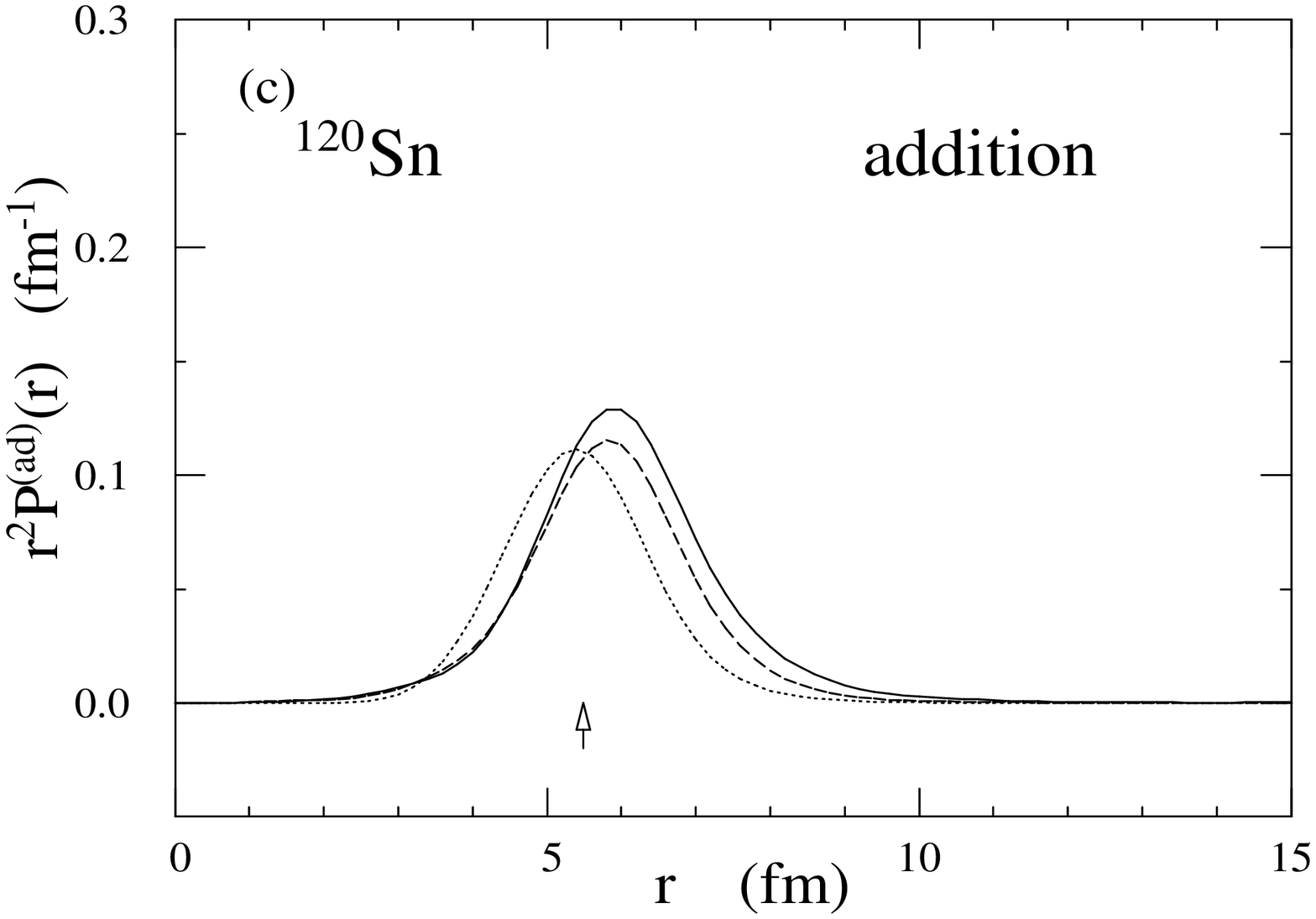}
\includegraphics[scale=0.4,angle=270]{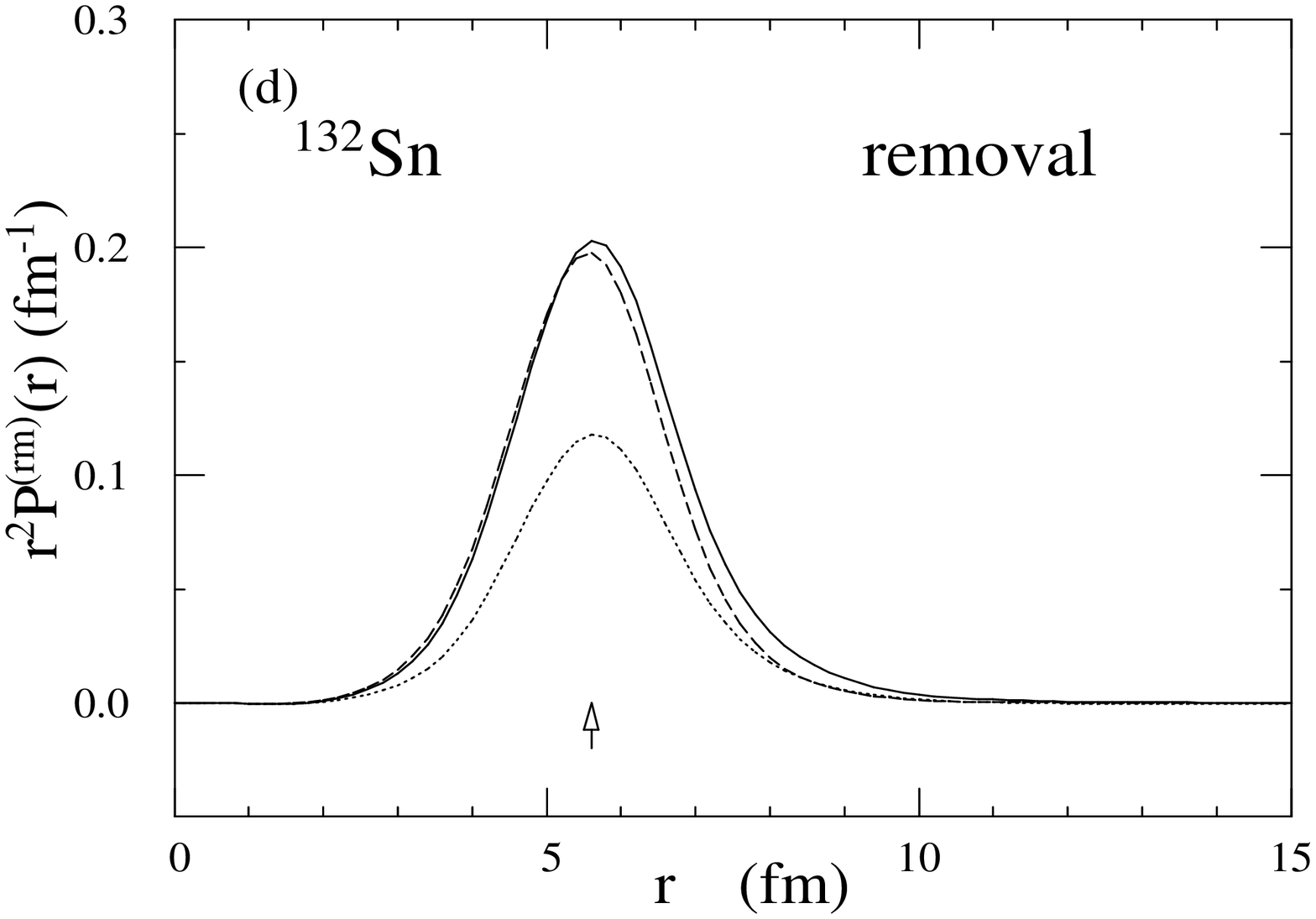}
\caption{\label{fig:transitionint} 
Effects of the pairing residual interactions on 
the neutron  pair-addition transition density $r^2P^{({\rm ad})}(r)$ 
feeding the $2_1^+$ states, calculated for the isotopes 
(a) $^{142}$Sn, (b) $^{134}$Sn, and (c) $^{120}$Sn. 
The solid and the dashed curves are the results obtained with the DDDI-bare'
and the volume pairing, respectively, while 
the unperturbed pair-addition transition density $r^2P^{({\rm ad})}_{{\rm unp}}(r)$ 
neglecting the residual interaction is also plotted with the dotted curve. The
arrow indicates the position of the half density radius $R_{1/2}$.
(d)The same as (a-c), but for 
the pair-removal transition density $r^2P^{({\rm rm})}(r)$ of the $2^+_1$ pair-removal 
mode in $^{132}$Sn.
}
\end{figure*}

\subsection{Sensitivity to the surface-enhanced pairing}\label{sec:sensitivity}

We shall now discuss the role of the residual pairing interaction and the
sensitivity of the quadrupole pair transfer to the enhancement
of the pairing interaction in the surface and the external
regions of a nucleus.
In Fig.\ref{fig:pdd-sn120-sn134} we compare the pair-addition strength 
function in $^{134}$Sn and $^{120}$Sn,
calculated with two  pairing interactions, the 
DDDI-bare' and the volume pairing, which have very different
density-dependence in the pairing interaction strength 
(cf. Fig.\ref{fig:DDDI}). In the case of $^{134}$Sn
we see a large difference in the quadrupole pair-addition strength
in $^{134}$Sn while  
in the case of $^{120}$Sn the difference is small.

A systematic comparison is made  in Fig.\ref{fig:pairstrength}
for the pair-addition and -removal 
strengths $B({\rm Pad}2)$ and $B({\rm Prm}2)$ 
of the $2_1^+$ mode calculated 
with the three kinds of the pairing interactions, including the mix pairing as well.
It is seen that the pair-addition strength exhibits
a sensitivity in all the nuclei to some extent, and the sensitivity
becomes prominent in $^{132-142}$Sn. The ratio of $B({\rm Pad}2)$ obtained with
the DDDI-bare' and that with the volume pairing is  
2.3, 2.8, 4.0, 3.5, 2.8 for $A=134$,136,138, 140, 142, respectively, while 
for $A <132$ it is in the range
1.4-2.0.  Note, on the other hand, that the pair-removal strength
$B({\rm Prm}2)$
is much less sensitive in all the isotopes. The difference
in $B({\rm Prm}2)$ for the different pairing interactions is negligible  
in many isotopes.

Let us analyze the mechanism of the sensitivity of
the pair-addition strength by looking into the
transition densities. 
We show in Fig.~\ref{fig:transitionint} (a)-(c)
the transition density $P^{({\rm ad})}(r)$ of the pair-addition mode
calculated with different pairing interactions 
in the three representative Sn isotopes. 
In this figure we see an intimate relation between the degree of sensitivity and the
radial profile of the transition density. 
Namely a large sensitivity observed 
in $^{134}$Sn and $^{142}$Sn can be linked to
the pair-addition transition density extending  
far outside the nuclear surface, $r \gesim R_{1/2}+2$fm.
The sensitivity is small in $^{120}$Sn where 
the transition density is confined mostly within the surface 
region $r \lesim R_{1/2}+2$fm.

For more detailed discussion let us examine the values of 
the pair-addition transition density $P^{({\rm ad})}(r)$ in $^{134}$Sn
at $r=5.8, 7.8, 9.8$ fm,
corresponding approximately to the half density radius $r=R_{1/2}$ ($=5.70$ fm),
and $r=R_{1/2}+2$ fm, and $r=R_{1/2}+4$ fm. The ratio of 
$P^{({\rm ad})}(r)$ obtained with the DDDI-bare'  and that with the
 volume pairing is 1.45, 1.82 and 2.71 for
$r=5.8, 7.8, 9.8$ fm, respectively, indicating an increase of the
sensitivity 
as moving from the surface toward the outside. 

It is also useful to compare $P^{({\rm ad})}(r)$ with
the transition density $P^{({\rm ad})}_{{\rm unp}}(r)$ associated with the unperturbed 
 pair-addition transfer (i.e. the transition associated with
the lowest-energy two-quasiparticle excitation
obtained by neglecting the residual pairing
interaction), which is also shown
in Fig.\ref{fig:transitionint} with the dotted curve.
 The  ratio between 
the transition density $P^{({\rm ad})}(r)$ 
obtained with the DDDI-bare' (or the volume pairing) and
the unperturbed transition density $P^{({\rm ad})}_{{\rm unp}}(r)$ 
is a measure of the influence of the RPA
correlation (the configuration mixing) caused by the residual pairing interaction. 
In the case of the DDDI-bare',
the ratio $P^{({\rm ad})}(r)/P^{({\rm ad})}_{{\rm unp}}(r)$ reads
2.7 at the nuclear surface $r =R_{1/2}$, and
it increases to 3.8-6.2 at $r= (R_{1/2}+2 {\rm fm}) -
(R_{1/2} + 4 {\rm fm}) \approx 8 - 10 {\rm fm}$. 
This is in contrast to the volume
pairing causing much smaller ratio 
1.9-2.1-2.3 at the same positions. 
We see even stronger sensitivity to the density-dependence of
the pair interaction in $^{142}$Sn, for which the tail extends further outside.

We thus see that the strong sensitivity of the pair-addition 
transfer 
originates from the RPA correlation 
taking place in the surface and exterior regions in $^{134}$Sn and $^{142}$Sn. 
We recall the reader here that 
the three pairing interactions have interaction strengths different by
a factor of more than 2 (cf. Fig.\ref{fig:DDDI}) at the positions $r \gesim R_{1/2}+2$ fm
corresponding to 
low densities $\rho/\rho_0 \lesim 0.1$.

A contrasting case is $^{120}$Sn shown in
Fig.~\ref{fig:transitionint}(c), in which 
we see very little the RPA correlation effect and the
difference between the DDDI-bare' and the volume pairing interactions.
This is what we can expect from the discussions above 
as there is essentially no significant 
spatial extension of the pair-addition transition density 
in this tightly bound stable isotope, as seen in Fig.~\ref{fig:transitionint}(c).

In  Fig.~\ref{fig:transitionint}(d)  we show the transition density 
$P^{({\rm rm})}(r)$ of the
pair-removal mode in $^{132}$Sn. 
In this case
the pair-removal
amplitude is concentrated only in the vicinity and the inside of
the surface $r \lesim R_{1/2}+2$ fm, exhibiting a very different 
radial profile from that of the pair-addition mode 
in $^{134,142}$Sn (Fig.~\ref{fig:transitionint}(a) and (b)). 
It is also seen  that
the pair-removal transition density $P^{({\rm rm})}(r)$ does not depend
on which we adopt
the DDDI-bare' or the volume pairing.
We can interpret this insensitivity in terms of  Fig.~\ref{fig:DDDI}:
the effective pairing strengths of the three interactions
are comparable in size on average 
in the region $ |r - R_{1/2}| \lesim 2$ fm
where the transition strength is concentrated.

Finally we remark 
that the three pairing interactions give essentially
identical results for $E(2_1^+)$ and $B({\rm E}2)$ except for $A \ge 140$,
as is seen  in Fig.~\ref{fig:Exstrength} (a) and (b).
The excitation energy
and the $B({\rm E}2)$ 
of the $2_1^+$ state in superfluid nuclei is known to depend
 on the pairing correlation\cite{BM2}. However,
the results indicate that 
these quantities are not influenced very much by 
the density-dependence of the pairing interaction
as far as the average pairing gap is constrained to the same value.

\section{Conclusions}

We have investigated the neutron pair transfer modes with 
the quadrupole multipolarity feeding the 
low-lying quadrupole states
in the neutron-rich Sn isotopes by means
of the QRPA based 
on the Skyrme-Hartree-Fock-Bogoliubov mean-fields. 
As a model of the surface-enhanced paring, we used
the density-dependent delta interaction (DDDI) whose strength 
increases at low nucleon densities reflecting 
the large scattering length of the $^{1}S$ bare interaction 
in the free space.

The numerical calculations indicate that the lowest-lying quadrupole state
exhibits a sudden change of its character as the neutron number
increases across the magic number $N=82$, beyond which features of
weakly binding emerge in the neutron motion.
 The most noticeable feature is found in the pair-addition mode.
Namely the transition
density of the pair-addition mode feeding the $2_1^+$ state
extends spatially far outside the
nuclear surface with a long tail reaching $r\sim 12-13$ fm in
the isotopes $^{>132}$Sn.  Accordingly the strength of the quadrupole
neutron pair-addition transition is enhanced for $A>132$. 
The enhanced pair-addition transfer is a consequence of 
the RPA correlation caused by 
the pairing interaction acting strongly in the 
surface and external regions of a nucleus.

Comparing with other DDDI's having different density dependence, i.e.
the density-independent volume pairing
and the mixed pairing having an intermediate density dependence,  
it is found that the quadrupole pair-addition mode 
exhibits generally sensitivity to the increase of the
pairing interaction strength in
the surface and the external regions of a nucleus. 
The sensitivity of the pair-addition mode 
to the three interactions is about a factor of 1.4-2.0 
in the isotopes with $A<132$, and it is magnified 
to a factor of 2.0-4.0 in
the isotopes $A>132$. Therefore one can expect that the neutron 
pair-addition transfer feeding the $2_1^+$ state in 
the neighboring $A+2$ isotope,
especially those feeding the isotopes $A> 132$,
provides a probe to the enhancement of the pairing
interaction strength expected in the surface and the external regions. 
On the other hand,
the neutron pair-removal transfer feeding the $2_1^+$ state in
the $A-2$ isotope does not have the sensitivity, and neither
$B(E2)$ nor the quadrupole neutron strength $B(N2)$ does,
as far as the Sn isotopes analyzed in the present work are concerned.

\begin{acknowledgments}
One of the authors (M.M.) thanks Prof. M.~Kawai, Prof. M.~Ichimura  and
Prof. A.~Vitturi for useful discussions.
A part of the 
 numerical calculations were performed on the NEC SX-8 supercomputer
 systems at Yukawa Institute for Theoretical Physics, Kyoto University, 
and at Research
 Center for Nuclear Physics, Osaka University. The work was supported by
the Grant-in-Aid for Scientific Research 
(Nos. 20540259, 21105507 and 21340073) from the Japan
 Society for the Promotion of Science,   and also by
the JSPS Core-to-Core Program, International
Research Network for Exotic Femto Systems(EFES).

\end{acknowledgments}

\end{document}